\documentclass[prd,preprint,showpacs,eqsecnum]{revtex4}

\usepackage{amsfonts}
\usepackage{amssymb}
\usepackage{graphicx}
\usepackage{pstricks}
\usepackage{axodraw}

\newcommand{\be}{\begin{equation}}
\newcommand{\bea}{\begin{eqnarray}}
\newcommand{\ee}{\end{equation}}
\newcommand{\eea}{\end{eqnarray}}
\newcommand{\bpi}{\begin{picture}}
\newcommand{\bce}{\begin{center}}
\newcommand{\epi}{\end{picture}}
\newcommand{\ece}{\end{center}}

\newcommand{\ga}{\Gamma}
\newcommand{\gp}{\Gamma^{\rm P}}
\newcommand{\gf}{\Gamma^{\rm F}}

\newcommand{\de}{d^{-1}}
\newcommand{\di}{d_{\rm i}}

\newcommand{\Qpsm}{Q\hspace{-0.26cm}/\,'}
\newcommand{\Qsm}{Q\hspace{-0.26cm}/}
\newcommand{\ksm}{k\hspace{-0.24cm}/}
\newcommand{\qsm}{q\hspace{-0.22cm}/}

\def\g{{\rm I}\hspace{-0.07cm}\Gamma}

\def\r#1{(\ref{#1})}

\begin{document}

\title{Pinch Technique and the Batalin-Vilkovisky formalism}
\date{April 11, 2002}

\author{Daniele Binosi}
\author{Joannis Papavassiliou}
\affiliation{Departamento de F\'\i sica Te\'orica and IFIC, Centro Mixto, 
Universidad de Valencia-CSIC,
E-46100, Burjassot, Valencia, Spain}

\email{Daniele.Binosi@uv.es; Joannis.Papavassiliou@uv.es}

\begin{abstract}

In  this paper  we  take  the first  step  towards a  non-diagrammatic
formulation of the  Pinch Technique.  In particular we  proceed into a
systematic identification  of the parts  of the one-loop  and two-loop
Feynman  diagrams that are  exchanged during  the pinching  process in
terms of unphysical ghost Green's  functions; the latter appear in the
standard  Slavnov-Taylor Identity  satisfied  by the  tree-level  and
one-loop  three-gluon  vertex.   This  identification allows  for  the
consistent  generalization of  the  intrinsic Pinch  Technique to  two
loops, through  the collective treatment  of entire sets  of diagrams,
instead of the laborious  algebraic manipulation of individual graphs,
and sets  up the  stage for  the generalization of  the method  to all
orders.   We  show  that the task of  
comparing the  effective  Green's
functions  obtained by the  Pinch Technique  with those
computed in the Background Field Method Feynman gauge is significantly
facilitated  when employing  the powerful  quantization  framework of
Batalin and Vilkovisky.  This formalism allows for the derivation of a
set  of useful  non-linear  identities, which  express the  Background
Field Method Green's functions  in terms of the conventional (quantum)
ones  and auxiliary  Green's functions involving
the background source and the gluonic anti-field;  
these  latter Green's functions are  subsequently
related by  means of a Schwinger-Dyson  type of equation  to the ghost
Green's  functions  appearing  in  the  aforementioned  Slavnov-Taylor
Identity.

\end{abstract}


\preprint{FTUV-02-0411}
\preprint{IFIC-02-09}

\maketitle

\section{Introduction}

The  Pinch Technique  (PT)
\cite{Cornwall:1982zr,Cornwall:1989gv,Papavassiliou:1990zd} is  a
diagrammatic  method which exploits  the  underlying  symmetries 
encoded  in  a  {\it  physical} amplitude  such  as  an  $S$-matrix  element,
in  order  to  construct effective   Green's   functions    with   special  
properties.    The aforementioned symmetries,  even though they are  always
present, they are  usually concealed by  the gauge-fixing  procedure.  The  PT
makes them  manifest by means  of a  fixed algorithm,  which does  {\it not}
depend on  the gauge-fixing scheme  one uses in order  to quantize the theory,
{\it i.e.}, regardless of the  set of Feynman rules used when writing down  the 
$S$-matrix  element.   The  PT  exploits  the elementary  Ward
Identities (WIs) 
triggered  by the  longitudinal  momenta appearing  inside  Feynman  diagrams  
in  order  to  enforce massive cancellations. The realization of these
cancellations  mixes non-trivially contributions stemming from diagrams of
different  kinematic nature (propagators, vertices, boxes). Thus,      a  given
physical  amplitude  is reorganized into sub-amplitudes,   which  have   the 
same   kinematic   properties  as conventional $n$-point functions and, in 
addition, are  endowed with  desirable physical  properties.  Most importantly,
at  one- and two-loop  order they are independent  of the gauge-fixing 
parameter,  satisfy  naive  (ghost-free) tree-level WIs
instead of the usual Slavnov-Taylor  identities (STIs) 
\cite{Slavnov:1972fg,Taylor:1971ff},
and contain only physical thresholds
\cite{Papavassiliou:1995fq,Papavassiliou:1996gs}.

It is clear  by now that an intimate connection  exists between the PT and 
the  Background  Field  Method  (BFM)    
\cite{Dewitt:1967ub, Honerkamp:1972fd,Kallosh:1974yh,
Kluberg-Stern:1975xv,Arefeva:1975jv,'tHooft:1975vy, 
Abbott:1981hw, Weinberg:1980wa,Shore:1981mj,
Abbott:1983zw,Hart:1983jy}.
The  BFM  is  a  special gauge-fixing 
procedure, implemented  at the  level of  the generating functional.  In 
particular, it preserves  the symmetry of  the action under ordinary  gauge
transformations  with respect to  the background (classical) gauge field
$\widehat{A}_{\mu}$,  while the quantum gauge fields $A_{\mu}$ appearing in
the  loops transform  homogeneously under  the gauge  group,  {\it i.e.},
as ordinary matter  fields which happened  
to be assigned to  the adjoint representation 
\cite{Weinberg:kr}.   As a  result  of the  background gauge symmetry, the BFM
$n$-point  functions  $\langle 0  | T \left[ \widehat{A}_{\mu_1}(x_1)
\widehat{A}_{\mu_2}(x_2)\dots  \widehat{A}_{\mu_n}(x_n) \right] |0 \rangle$
satisfy naive QED-like WIs, but (unlike QED) depend  explicitly on
the quantum gauge-fixing parameter
$\xi_Q$ used  to define the tree-level  propagators of  the 
quantum gluons.   It  turns out that at one-loop  order, both in QCD  and in
the Electroweak  sector   of  the  Standard   Model,  the  
 gauge-fixing parameter-independent
effective $n$-point functions constructed by means of the PT (starting from 
any   gauge-fixing  scheme)  coincide   with  the  corresponding background 
$n$-point functions when  the latter  are computed  at the special   value 
$\xi_Q =1$   (BFM Feynman  gauge) 
\cite{Denner:1994nn, Hashimoto:1994ct, Pilaftsis:1997fh} . 
As   was  shown   
in detail in \cite{Papavassiliou:1999az,Papavassiliou:1999bb},  
this correspondence  {\it persists} at two loops
in  the case  of QCD.  

One of the most pressing questions  in this context is whether one can
extend the  PT algorithm  to all orders  in perturbation  theory, thus
achieving the systematic construction of effective $n$-point functions
displaying the aforementioned  characteristic features.  To accomplish
this  it  is clear  that  one needs  to  go  beyond the  diagrammatic
manipulations  employed  until  now,  and  resort  to  a  more  formal
procedure.  Indeed, one disadvantage of the PT method is the fact that
the  constructions   rely  heavily  on   algebraic  operations  inside
individual  Feynman  graphs.   Even  though these  operations  proceed
according to  well-defined guiding principles which  have been spelled
out in  various occasions in  the existing literature, any  attempt to
apply them to higher orders would constitute an operationally hopeless
task.   But even  if  the  resulting re-shuffling  of  terms among  the
Feynman  graphs  would  eventually  lead  to  a  well-defined  answer,
additional effort  would be required  in order to compare  this unique
answer to the BFM $n$-point  functions, and to verify whether the
correspondence mentioned above persists to all orders.

To  ameliorate this  situation, in  this paper  we take  a  first step
towards  a  non-diagrammatic  formulation  of the  PT  procedure.   In
particular we proceed into a systematic identification of the parts of
the one-loop  and two-loop Feynman  diagrams that are  shuffled around
during the pinching process in terms of well-defined field-theoretical
objects, namely  the ghost Green's  functions which appear in  the STIs
satisfied  by the  tree-level and  one-loop three-gluon  vertex 
\cite{Ball:1980ax}. This
constitutes an important step because  it enables one to go beyond the
current diagrammatic implementation of the pinching procedure by means
of tree-level WIs appearing  in individual graphs, allowing instead the
collective treatment of entire sets of diagrams, and sets up the stage
for the generalization of the method to all orders \cite{DBJP2}. 
Thus, at least at
one- and two-loops, the final  PT answer for a given effective Green's
function is obtained from the  original Green's function by adding (or
subtracting) a  well-defined set of contributions  identified when the
relevant  STIs have  been triggered  {\it inside}  the  Green's function
under consideration.

The conventional derivation of the STIs using the Becchi-Rouet-Stora-Tyutin
(BRST) transformations \cite{Becchi:1975md,Becchi:1976nq,Tyutin:1975qk}
and  the definition  of the  building  blocks in  terms of  unphysical
ghost-Green's   functions   is   in   itself   a   text-book   exercise
\cite{Pascual:zb}.  But in addition,  we will carry out the derivation
of the very same STIs  using the Batalin-Vilkovisky (BV) formalism
\cite{Batalin:1983jr,Batalin:1977pb}. In
particular, the STIs  written in the context of the  BV are realized by
means of auxiliary unphysical  Green's functions, which involve ghosts
and anti-fields;  the latter are  characteristic of the  BV formalism,
and do not appear in the conventional formulation of the gauge theory.
Of  course,  since the  STI  in  both  formulation involves  the  same
original Green's  function, namely  it is the  STI of  the three-gluon
vertex,  the  building  blocks   appearing  in  the  two
formulations-- 
conventional and BV-- must be related. It turns out that this indeed
the case,  as we  will see in  detail in  Section~\ref{sec:one}. 
The  reason for
going through this  exercise is because thusly one  may take advantage
of  an important  ingredient furnished  by the  BV  formulation, which
facilitates significantly the comparison  of the PT results with those
of the BFM.  Specifically using  the formulation of the BFM within the
BV formalism,  one can derive non-trivial identities  relating the BFM
$n$-point  functions  to   the  corresponding  conventional  $n$-point
functions  in the covariant  renormalizable gauges,  to all  orders in
perturbation   theory.    These  identities,   which   we  will   call
Background-Quantum identities (BQIs) in what follows,  
have been derived for the first time in the context 
of the Standard Model in  \cite{Gambino:1999ai,Grassi:2001zz}.  
The quantities appearing in  these BQIs are Green's functions involving
anti-fields   and   background   sources,   introduced   in   the   BFM
formulation.  It  turns  out  that  the  auxiliary  Green's  functions
appearing in  the STIs and those  appearing in the BQIs,  are related by
simple expressions, a fact which allows for a direct comparison of the
PT and BFM Green's functions.   Notice that the BV formalism furnishes
exact  Feynman   rules  for  the  perturbative   construction  of  all
aforementioned  unphysical, auxiliary  Green's function,  appearing in
the STIs and the BQIs.

It is conceptually very  important to emphasize the logical succession
of the steps  involved in this entire construction:  One begins with a
massless   Yang-Mills  theory,   such  as   QCD,  formulated   in  the
conventional way, {\it i.e.}, with a linear covariant gauge-fixing term
of   the   form   $\frac1{2\xi}\left(\partial^\mu   A^a_\mu\right)^2$,
together  with  the  corresponding  ghost-sector,  introduced  by  the
standard Faddeev-Popov  construction; at this stage  this theory knows
nothing about neither  the BFM nor the anti-fields  appearing in the
BV formulation.   Exploiting only  the STIs, derived  by virtue  of the
BRST  symmetry and  formulated  in the  language  of the  conventional
theory,  {\it i.e.},  expressed  solely in  terms of  objects definable
within this  theory, one can reach  after a well-defined  set of steps
the PT answer.  The most  expeditious way for comparing this answer to
the corresponding  BFM Green's function is the  following: one derives
the  aforementioned  STIs  using  the  BV formalism,  {\it  i.e.}   one
translates the  STIs from the normal  language to the  BV language; the
reason is that  thusly one can exploit the  identities -- derivable in
the BV  language -- relating the  BFM Green's functions  to the normal
ones.

The paper  is organized as follows:  In Section~\ref{sec:one} 
we  present a brief
introduction  to the  BV formalism,  providing the  minimum  amount of
information  needed  for establishing  notation  and  arriving at  the
relevant  generating  functional.   In Section~ \ref{sec:two}
 we  derive  the
necessary ingredients following  standard manipulations: In particular
we  derive the  STIs within  the BV,  as well  as the  BQIs  for various
cases.  In addition,  we  derive a  Schwinger-Dyson  type of  identity
relating the building  blocks appearing in the STIs  to those appearing
in the BQIs; to the best of our knowledge this relation appears for the
first  time  in  the literature.   In  Section~\ref{sec:three}  
we review  the  PT
construction, and put to work the formalism derived above. Even though
the PT part  is standard, this section provides  a distilled review of
the PT method, and serves  as a simple testing-ground for establishing
the desired  connections between the  two formalisms. 
In Section~\ref{sec:four} we
present the  two-loop construction, where  the connections established
are further scrutinized, within a far more complex context. 
In  Section~\ref{secIP} we  present an
entirely new result,  even from the point of  view of conventional PT,
namely the  two-loop generalization of the  intrinsic PT construction.
In particular, we  will show how the judicious  organization of entire
sets of  two-loop diagrams, together with  the use of the  STI for the
one-loop three-gluon vertex,  leads to the PT answer  for the two-loop
effective gluon self-energy. Finally, in Section~\ref{sec:five} 
we present our conclusions. 

\section{\label{sec:one}The Batalin-Vilkovisky formalism}

In this section we will briefly review the most salient features of the BV
formalism \cite{Batalin:1983jr,Batalin:1977pb}, concentrating to its 
application to the case of massless Yang-Mills theories. 

The (gauge fixed) Yang-Mills Lagrangian density will be given by
\be
{\cal L}_{\rm YM}={\cal L}_{\rm I} +{\cal L}_{\rm GF}+{\cal L}_{\rm FPG},
\label{YM}
\ee
with ${\cal L}_{\rm I}$ the usual gauge invariant $SU(N)$ 
Yang-Mills Lagrangian, 
\be
{\cal L}_{\rm I} = - \frac{1}{4} F_{\mu\nu}^a F^{a\,\mu\nu}
+\bar\psi\left(iD\hspace{-0.26cm}/-m\right)\psi, 
\ee
where 
\be
F_{\mu\nu}^a = 
\partial_{\mu} A_{\nu}^a - \partial_{\nu} A_{\mu}^a + 
 g f^{abc} A_{\mu}^b  A_{\nu}^c,
\ee
$g$ is the gauge coupling, and $D_\mu$ is the covariant derivative
defined as 
\be
D_\mu=\partial_\mu-igT^aA^a_\mu.
\ee
The covariant gauge fixing and Faddeev-Popov term 
${\cal L}_{\rm GF}+{\cal L}_{\rm FPG}$ will be chosen to have the form
\be
{\cal L}_{\rm GF}+{\cal L}_{\rm FPG}
= -\frac12\xi \left(B^a\right)^2+
B^a\partial^\mu
A^a_\mu
-\bar c^a\partial^\mu\left(\partial_\mu
c^a-g f^{abc}A^b_\mu c^c\right).
\ee
The $B^a$ are auxiliary, non-dynamical fields, 
since they have a quadratic term without derivatives 
(and as such they are not propagating). They  
represent the so-called Nakanishi-Lautrup Lagrange multiplier for the 
gauge condition, and they are usually eliminated through the corresponding
Gaussian integration in the path integral, 
giving rise to the usual gauge fixing term 
\be
{\cal L}_{\rm GF}=\frac1{2\xi}\left(\partial^\mu 
A^a_\mu\right)^2.
\ee

The starting point of the BV formalism is the introduction of an {\it external}
field -- called anti-field --
$\Phi^{*,n}$ for each field  $\Phi^n$ appearing in the Lagrangian. 
In particular, here $\Phi^n$ represent generically any of the fields 
$A^a_\mu,c^a,\bar c^a,\psi,\bar\psi$ and $B^a$
appearing in Eq.\r{YM}. The anti-fields $\Phi^{*,n}$ will carry the 
same Bose/Fermi
statistic of the corresponding field $\Phi^n$ and a ghost number such that
\be
gh\,\{\Phi^{*,n}\}=-gh\,\{\Phi^{n}\}-1.
\ee
Thus, since the
ghost number is equal to $1$ for the ghost fields 
$c^a$, to $-1$ for the anti-ghost fields $\bar c^a$, 
and zero for the other fields, one has the assignment  
\be
gh\,\{A^{*,a}_\mu,c^{*,a},\bar c^{*,a},\psi^*,\bar\psi^*\}=\{-1,-2,0,-1,-1\}.
\ee
The original gauge invariant Lagrangian is then supplemented with a term 
coupling to the anti-fields $\Phi^{*,n}$ with the BRST variation of $\Phi^n$, 
giving the modified Lagrangian
\bea
{\cal L}_{\rm BV}&=&{\cal L}_{\rm I}+{\cal L}_{\rm BRST} \nonumber \\
&=&{\cal L}_{\rm I}+\sum_n\Phi^{*,n}s\Phi^n,
\eea
with $s$ the BRST operator, and
\bea
sA^a_\mu= \partial_\mu c^a-g f^{abc}A^b_\mu c^c, &\qquad& 
sc^a=-\frac12 g f^{abc}c^bc^c, \nonumber \\
s\psi=ig c^aT^a\psi, &\qquad& s\bar\psi=-ig \bar\psi T^ac^a, \nonumber \\
s\bar c^a=B^a, &\qquad& sB^a=0. 
\eea
The action $\g^{(0)}[\Phi,\Phi^*]$ built up from the new Lagrangian 
${\cal L}_{\rm BV}$, will 
then satisfy the {\it master equation}
\be
\int\!d^4x\left[\frac{\delta\g^{(0)}}{\delta\Phi^{*,n}}\frac{\delta\g^{(0)}}
{\delta\Phi^n}\right]=0,
\label{mecl}
\ee
which is just a consequence of the BRST invariance of the action and of the 
nilpotency of the BRST operator. 

Since the  anti-fields are  external fields, we  must constrain  them to
suitable  values  before we  can  use  the  action $\g^{(0)}$  in  the
calculation of $S$-matrix elements. To this purpose one introduces an
arbitrary fermionic functional $\Psi[\Phi]$ (with
$gh\,\{\Psi[\Phi]\}=-1$) such that            
\be
\Phi^{*,n}=\frac{\delta\Psi[\Phi]}{\delta\Phi^n}.  
\ee
Then the action
becomes                                                            
\bea
\g^{(0)}[\Phi,\delta\Psi/\delta\Phi]&=&\g^{(0)}[\Phi]+(s\Phi^n)
\frac{\delta\Psi[\Phi]}{\delta\Phi^n}\nonumber   \\
&=&\g^{(0)}[\Phi]+s\Psi[\Phi],
\eea 
{\it i.e.}, it is equivalent to the
gauge fixed action of the  Yang-Mills theory under scrutiny, since we
can choose the fermionic functional $\Psi$ to satisfy
\be
s\Psi[\Phi]=\int\!d^4x\left({\cal L}_{\rm GF}+{\cal L}_{\rm
FPG}\right).  
\ee
The fermionic functional $\Psi$ is often referred to
as the gauge fixing fermion. 

Moreover, the auxiliary fields $B^a$ and the anti-ghost anti-fields 
$\bar c^{*,a}$ have linear BRST transformations, so that they form a 
so called {\it trivial pair} \cite{Barnich:2000zw}: they enter, 
together with their anti-fields, bilinearly in the action 
\be
\g^{(0)}[\Phi,\Phi^*]=\g^{(0)}_{\rm min}[A^a_\mu,c^a,A^{*,a}_\mu,c^{*,a}]-
B^a\bar c^{*,a}.
\ee
The last term has no effect on the master equation, which will be in fact
satisfied by the {\it minimal} action $\g^{(0)}_{\rm min}$ alone. 
In what follows we will restrict our considerations to the minimal 
action (which depends on the {\it minimal variables}  
${A^a_\mu,c^a,A^{*,a}_\mu,c^{*,a}}$), 
dropping the corresponding subscript. 

It is well known that the BRST symmetry is crucial for providing the
unitarity of the $S$-matrix and the gauge independence of physical
observables; thus it must be implemented in the theory to all orders,
not only at the classical level. This is provided by establishing the
quantum corrected version of Eq.\r{mecl}, in the form of the STI functional
\bea
{\cal S}(\g)[\Phi,\Phi^*]&=&\int\!d^4x\left[
\frac{\delta\g}{\delta\Phi^{*,n}}\frac{\delta\g}
{\delta\Phi^n}\right] \nonumber \\
&=&\int\!d^4x  \left[  
\frac{\delta \g}{\delta A^{*,a}_\mu}  \frac{\delta\g}{\delta A^a_\mu} 
+\frac{\delta \g}{\delta c^{*,a}}  \frac{\delta\g}{\delta c^a} 
+\frac{\delta \g}{\delta \psi^{*}}  \frac{\delta\g}{\delta \bar\psi}
+\frac{\delta \g}{\delta \psi}  \frac{\delta\g}{\delta \bar\psi^*}
\right] \nonumber \\
& = &  0,
\label{mequ}
\eea
where $\g[\Phi,\Phi^*]$ is now the effective action. Eq.\r{mequ} gives
rise to the complete set of non linear STIs at all orders in the perturbative
theory, via the repeated 
application of functional differentiation. Notice that $gh\,\{{\cal
S}(\g)\}=+1$ and that Green's functions with non-zero ghost charge
vanish, since it is a conserved quantity. This implies that
for
getting non-zero identities it is necessary to differentiate the
expression \r{mequ} with respect to one ghost field (ghost charge~+1)
or with respect to two ghost fields and one anti-field (ghost charge
$+2-1=+1$ again). For example, for deriving the STI satisfied by the
three-gluon vertex, one has to differentiate Eq.\r{mequ} with respect
to two gluon fields and one ghost field (see Section \ref{subSTI} below).

A technical remark is in order here. Recall that we have chosen to
work with the minimal generating functional $\g$, from which the
trivial pair $({B^a,\bar c^{*,a}})$ has been removed 
\cite{Barnich:2000zw,Barnich:2000yx}. In the case of a linear gauge
fixing as the one at hands, this is equivalent to working with the
``reduced'' functional $\g$, defined by subtracting from the complete
genereting functional $\g^{\scriptscriptstyle{\rm C}}$ the local term
$\int\!d^4x\ {\cal L}_{\rm GF}$ corresponding to the gauge-fixing part
of the Lagrangian. One  
should then  keep in mind that the Green's functions
generated by the minimal effective action $\g$ or the complete one
$\g^{\scriptscriptstyle{\rm C}}$ are not equal \cite{Gambino:1999ai}. 
At tree-level, one has for example that
\bea
\g_{A^a_\mu A^b_\nu}^{(0)}(q)&=&\g^{{\rm C}\,(0)}_{A^a_\mu
A^b_\nu}(q)+\frac1\xi q_\mu q_\nu \nonumber \\
&=&-i\delta^{ab}q^2P_{\mu\nu}(q),
\label{uns}
\eea  
where $P_{\mu\nu}=g_{\mu\nu}-q_\mu q_\nu/q^2$ is the dimensionless
transverse projector; at higher orders the difference depends only on
the renormalization of the gluon field and of the gauge parameter (and,
as such, is immaterial for our purposes).

Another important ingredient of the construction we carry out in what follows 
is to write down the STI functional in the BFM. 
For doing this we introduce a classical vector field $\Omega_\mu^a$ which 
carries the
same quantum numbers as the gluon but ghost charge~$+1$. We then 
implement the equations of motion of 
the background fields at the quantum level 
by extending the BRST symmetry to them through the equations
\be
s\widehat A^a_\mu=\Omega^a_\mu, \qquad s\Omega^a_\mu=0.
\ee
Finally, in order to control the dependence of the Green's functions on 
the background fields, we modify the STI functional of Eq.\r{mequ} as 
\cite{Grassi:1999tp,Grassi:2001zz}
\be
{\cal S}'(\g')[\Phi,\Phi^*]={\cal S}(\g')[\Phi,\Phi^*]+\Omega^a_\mu
\left(\frac{\delta\g}{\delta \widehat{A}^a_\mu} - 
\frac{\delta\g}{\delta A^a_\mu}\right),
\label{mequbfm}
\ee
where $\g'$ denotes the effective action that depends on the background 
sources  $\Omega^a_\mu$, and ${\cal S}(\g')[\Phi,\Phi^*]$ is the STI 
functional of Eq.\r{mequ}. Differentiation of the STI functional 
Eq.\r{mequbfm} with respect to the background source and background or
quantum 
fields, will then relate 1PI functions involving background fields with 
the ones 
involving quantum fields (see~Section~\ref{subBQI} below).

The final ingredient we need to know for the actual computation of STIs are 
the coupling of the anti-fields and background sources to the other
fields of the  
theory. These are controlled by the Lagrangians
\bea
{\cal L}_{\rm BRST} &=&
A^{*,a}_{\mu} \left[\partial_{\mu} c^{a} - g f^{abc}  
\left(A^b_\mu+\widehat A^b_\mu\right) c^{c}\right] 
-\frac{1}{2} g f^{abc} c^{*,a}c^{b} c^{c} 
+ig \left(\bar\psi^*c^aT^a\psi\right) +\, {\rm h.c.}, \nonumber \\
{\cal L}_{\Omega}&=&
\Omega^{a}_{\mu} \left[\partial_{\mu} \bar c^{a} - g f^{abc}  
\left(A^b_\mu+\widehat A^b_\mu\right) \bar c^{c}\right],
\eea 
from which the necessary Feynman rules can be derived. 
Notice that the Feynman rules for the vertices 
involving the background sources $\Omega^a$ are 
the same as the ones involving the 
anti-fields $A^{*,a}$ provided that we trade  
the ghost fields for anti-ghost fields. 

\section{\label{sec:two} The basic ingredients}

After having  reviewed the BV formalism  as it applies to  the case of
mass-less  Yang-Mills theories, we  next proceed  to derive  the basic
ingredients  needed for the  PT construction.   In particular  we will
focus on two aspects: ({\it i}) the  derivation of the STI for the off-shell
three-gluon vertex $\g_{A_\alpha A_\mu  A_\beta}(q_1,q_2,q_3)$; 
as we will see this
STI  is of  central  importance for  the  intrinsic PT  method, to  be
presented in  Section~\ref{secIP}.  Of  course the aforementioned STI  is known
since  a long-time  in  the  context of  the  standard formulation  of
non-Abelian  gauge theories following  the Faddeev-Popov  Ansatz
\cite{Ball:1980ax,Pascual:zb}; here
however we want to relate in a {\it manifest} way the pieces appearing
in it (ghost Green's  functions) with well-defined quantities emerging
in  the  BV  formalism,  {\it  i.e.}  auxiliary  (unphysical)  Green's
functions involving ghost fields and gauge bosons
anti-fields.  ({\it ii}) the derivation
of the BQIs relating the background and quantum two-~three- and
four-point functions.   These identities furnish  non-linear relations
between   the  two   kinds   of  Green's   functions  and   facilitate
significantly the eventual comparison between the effective PT Green's
functions and the BFM Green's functions, computed at $\xi_Q = 1$.  The
crucial point  is that the conventional Green's  functions are related
to the BFM ones by means of  the same type of building blocks as those
that appear in the STI of the three-gluon vertex, derived in
({\it i}), namely  auxiliary, unphysical Green's functions.  Even though the
set of such auxiliary Green's  functions appearing in ({\it i}) is different
from  that appearing  in ({\it ii}),  since the  former involves
ghost fields and gauge boson 
anti-fields, whereas  the latter gauge boson 
background sources  and anti-fields, it
turns  out  that  the  two   sets  are  related  by  a  rather  simple
Schwinger-Dyson-type of relation, which  we present here for the first
time,  in Eq.(\ref{gpert1}). This  relation constitutes  a non-trivial
ingredient, bound to play a  central role in the generalization of the
intrinsic PT to  all orders \cite{DBJP2}, 
and constitutes a  central result of this section.

\subsection{\label{subSTI} Slavnov-Taylor Identity 
for the three-gluon vertex}

The standard text-book derivation of the three-gluon vertex STI 
starts from the trivial identity \cite{Pascual:zb}
\be
\langle 0 | T \left[ A_{\mu}^m(x) 
\bar{c}^b(y)[\partial^{\nu}A_{\nu}^c(z) ]\right] |0 \rangle = 0,
\ee
which is re-expressed in terms of the BRST-transformed
fields, making also use of the equal-time commutation relation
of the fields. 
The quantity which appears naturally when following
this procedure and Fourier-transforming the identity into
momentum space, is  (from now on we assume that 
all momenta appearing in a given Green's function are entering, 
{\it i.e.}, $q_1+q_2+q_3=0$ in the case at hands)
\be
L^{abc}_{\alpha\beta}(q_1,q_2,q_3) \equiv 
\int d^4x\, d^4y\, e^{-iq_1 x} e^{-iq_3 y} f^{aem}
\langle 0  | T \left[ A_{\alpha}^e(x) c^m(x) \bar{c}^c(y)  
A^{b}_{\beta}(0) \right] |0 \rangle, 
\ee
which is written in the form
\be
L^{abc}_{\alpha\beta}(q_1,q_2,q_3)  \equiv
\bigg[H^{b'c'a}_{\alpha\beta'}(q_2,q_3,q_1) 
+ \Sigma^{ad}_{\alpha}(q_1) D^{de}(q_1) 
G_{\beta'}^{b'c'e}(q_2,q_3,q_1)\bigg]
 D^{cc'}(q_3) \Delta^{b'b \,\beta'}_{\beta}(q_2),
\label{DEFL}
\ee
where we  
define the (full) ghost and gluon propagators (in the Feynman gauge)
as follows 
\bea
D(p) & = & \frac{i}{p^2-iL(p)}, \nonumber\\
\Delta_{\mu\nu}(q) & = & -i\left[\Delta(q^2)
P_{\mu\nu}(q)+\frac{q_\mu q_\nu}{q^4}\right], \qquad
\Delta(q^2)=\frac1{q^2+i\Pi(q^2)}, \nonumber\\
\Delta_{\mu\nu}^{(0)}(q) & = & g_{\mu\nu}d(q) , \qquad d(q)=-iq^{-2} .
\eea
The scalar quantities 
$L(p)$ and $\Pi(q^2)$ represent respectively the ghost and gluon 
self-energies.
The functions $\Sigma^{ad}_{\alpha}(q_1)$ and
$G_{\beta'}^{b'c'e}(q_2,q_3,q_1)$ are defined by means
of the quantities 
\bea
N_{\mu}^{ab}(p) &\equiv&
\int d^4x \,e^{-ipx} \,f^{amn}
\langle 0  | T \left[ A_{\mu}^m(x) c^n(x) \bar{c}^b(0) 
 \right] |0 \rangle, \nonumber\\ 
M_{\mu}^{abc}(q_1,q_2,q_3) &\equiv&
\int d^4x\, d^4y \,e^{-iq_3 x} e^{-iq_2 y}
\langle 0  | T \left[ c^c(x) \bar{c}^b(y)  A_{\mu}^a (0)
 \right] |0 \rangle, 
\eea
as follows:
\bea
g N_{\mu}^{ab}(p) &\equiv&
-\Sigma_{\mu}^{ac}(p) D^{cb}(p),
\nonumber\\
M_{\mu}^{abc}(q_1,q_2,q_3)&\equiv&
g G_{a'b'c'}^{\mu'}(q_1,q_2,q_3) \Delta_{\mu'\mu}^{a'a}(q_1) 
D^{b'b}(q_2)D^{c'c}(q_3). 
\eea

Notice that (after eliminating the dependence on one momentum, using
the constraint due to momentum conservation)  
the Green's functions $H$ and $\Sigma$ have the following 
diagrammatic definition
\be
\bpi(0,160)(75,-120)

\PhotonArc(40,0)(25,0,180){-1.5}{8.5}
\DashCArc(40,0)(25,180,360){1}
\DashArrowLine(95,-30)(70,-5){1}
\Photon(70,5)(95,30){1.5}{4}
\GCirc(65,0){12}{0.8}
\GCirc(40,23.5){12}{0.8}
\GCirc(40,-25){12}{0.8}
\DashArrowLine(19.2,-13.7)(18.2,-12.3){1}
\DashArrowLine(57.2,-17.8)(56.2,-18.8){1}
\Vertex(15,0){1.8}

\Text(5,-1)[r]{$H_{\alpha\beta}^{(n)}(q_1,q_2)\,=$}
\Text(66,0)[c]{$\scriptstyle{{\cal K}^{(n_3)}_{\nu\beta}}$}
\Text(40,-25)[c]{$\scriptstyle{D^{(n_1)}}$}
\Text(40,23.5)[c]{$\scriptstyle{\Delta^{(n_2)}_{\mu\nu}}$}
\Text(20,0)[l]{\footnotesize{${q_{1\alpha}}$}}
\Text(150,20)[l]{{$n=n_1+n_2+n_3+1$}}
\Text(150,0)[l]{$n_1,\,n_2\ge0 \Leftrightarrow n_3\ge1$}
\Text(150,-30)[l]{${\cal K}^{(0)}_{\nu\beta}\,=$}
\Text(98,30)[l]{\footnotesize{${q_{2\beta}}$}}

\Photon(190,-25)(220,-25){1.5}{6}
\DashArrowLine(220,-35)(190,-35){1}

\Text(190,-20)[r]{$\scriptstyle{\nu}$}
\Text(220,-20)[l]{$\scriptstyle{\beta}$}

\PhotonArc(40,-90)(25,0,180){-1.5}{8.5}
\DashCArc(40,-90)(25,180,360){1}
\DashArrowLine(19.2,-103.7)(18.2,-102.3){1}
\DashArrowLine(57.2,-107.8)(56.2,-108.8){1}
\DashArrowLine(105,-90)(75,-90){1}

\GCirc(65,-90){12}{0.8}
\GCirc(40,-66.5){12}{0.8}
\GCirc(40,-115){12}{0.8}
\Vertex(15,-90){1.8}

\Text(0,-90)[r]{$\Sigma_{\alpha}^{(n)}(q_1)\,=$}
\Text(66,-90)[c]{$\scriptstyle{G^{( n_3)}_{\nu}}$}
\Text(40,-115)[c]{$\scriptstyle{D^{(n_1)}}$}
\Text(40,-66.5)[c]{$\scriptstyle{\Delta^{(n_2)}_{\mu\nu}}$}
\Text(20,-90)[l]{\footnotesize{$q_{1\alpha}$}}
\Text(150,-90)[l]{{$n=n_1+n_2+n_3+1$}}

\epi
\label{pertexp}
\ee
which at tree-level implies \cite{Pascual:zb}
\be
\bpi(0,50)(55,-20)

\PhotonArc(40,0)(25,40,180){-1.5}{6.5}
\DashCArc(40,0)(25,180,320){1}
\DashArrowLine(40.5,-25)(39.5,-25){1}
\Vertex(15,0){1.8}

\Text(5,-1)[r]{$H_{\alpha\beta}^{(0)}(q_1,q_2)\,=$}
\Text(20,0)[l]{\footnotesize{$q_{1\alpha}$}}
\Text(62,18)[l]{\footnotesize{$q_{2\beta}$}}
\Text(125,-1)[r]{$=\,-igg_{\alpha\beta}$}

\epi
\ee

Clearly, by definition in Eq.(\ref{DEFL}), $H_{\alpha\beta}(q_1,q_2)$ 
corresponds the one-particle irreducible part of
$L_{\alpha\beta}(q_1,q_2)$, {\it i.e.}, graphically 

\be
\bpi(0,80)(150,-40)

\PhotonArc(40,0)(25,0,180){-1.5}{8.5}
\DashCArc(40,0)(25,180,360){1}
\PhotonArc(90,0)(25,30,180){-1.5}{8.5}
\DashCArc(90,0)(25,180,330){1}
\DashArrowLine(19.2,-13.7)(18.2,-12.3){1}
\DashArrowLine(57.2,-17.8)(56.2,-18.8){1}
\DashArrowLine(107.7,-17.1)(106.8,-17.9){1}
\DashArrowLine(73.6,-18.8)(72.6,-17.8){1}
\Vertex(15,0){1.8}

\GCirc(90,23.5){12}{0.8}
\GCirc(90,-25){12}{0.8}
\GCirc(65,0){12}{0.8}
\GCirc(40,23.5){12}{0.8}
\GCirc(40,-25){12}{0.8}

\Text(0,0)[r]{$L_{\alpha\beta}(q_1,q_2)\,=$}
\Text(66,0)[c]{$\scriptstyle{{\cal K}^{( n_3)}_{\nu\rho}}$}
\Text(40,-25)[c]{$\scriptstyle{D^{(n_1)}}$}
\Text(40,23.5)[c]{$\scriptstyle{\Delta^{(n_2)}_{\mu\nu}}$}
\Text(90,-25)[c]{$\scriptstyle{D^{(n_4)}}$}
\Text(90,23.5)[c]{$\scriptstyle{\Delta^{(n_5)}_{\rho\sigma}}$}
\Text(20,0)[l]{\footnotesize{$q_{1\alpha}$}}
\Text(115,12)[l]{\footnotesize{$q_{2\beta}$}}
\Text(145,0)[c]{$+$}

\PhotonArc(200,0)(25,0,180){-1.5}{8.5}
\DashCArc(200,0)(25,180,360){1}
\DashArrowLine(262.5,0)(225,0){1}
\DashArrowLine(300,0)(262.5,0){1}
\GCirc(225,0){12}{0.8}
\GCirc(200,23.5){12}{0.8}
\GCirc(200,-25){12}{0.8}
\GCirc(262.5,0){12}{0.8}
\PhotonArc(325,0)(25,30,180){-1.5}{8.5}
\DashCArc(325,0)(25,180,330){1}
\DashArrowLine(179.2,-13.7)(178.2,-12.3){1}
\DashArrowLine(217.2,-17.8)(216.2,-18.8){1}
\DashArrowLine(342.7,-17.1)(341.8,-17.9){1}
\DashArrowLine(308.6,-18.8)(307.6,-17.8){1}
\GCirc(300,0){12}{0.8}
\GCirc(325,23.5){12}{0.8}
\GCirc(325,-25){12}{0.8}
\Vertex(175,0){1.8}

\Text(227,0)[c]{$\scriptstyle{G^{(\tilde n_3)}_{\nu}}$}
\Text(301,0)[c]{$\scriptstyle{G^{(\tilde n_5)}_{\rho}}$}
\Text(201,-25)[c]{$\scriptstyle{D^{(\tilde n_1)}}$}
\Text(201,23.5)[c]{$\scriptstyle{\Delta^{(\tilde n_2)}_{\mu\nu}}$}
\Text(263.5,0)[c]{$\scriptstyle{D^{(\tilde n_4)}}$}
\Text(326,-25)[c]{$\scriptstyle{D^{(\tilde n_6)}}$}
\Text(326,23.5)[c]{$\scriptstyle{\Delta^{(\tilde n_7)}_{\rho\sigma}}$}
\Text(180,0)[l]{\footnotesize{$q_{1\alpha}$}}
\Text(350,12)[l]{\footnotesize{$q_{2\beta}$}}

\epi 
\ee

Notice the constraint on the values of $n_1$, $n_2$, and $n_3$, 
appearing in the definition of $H_{\alpha\beta}^{(n)}(q_1,q_2)$. 
Clearly, since $H_{\alpha\beta}^{(n)}(q_1,q_2)$ corresponds to an 
{\it amputated} vertex, $n_1$ and $n_2$ may differ from zero,
only iff $n_3 \geq 1$.  

After standard manipulations one arrives at the  
well-known STI \cite{Ball:1980ax}
\bea
q_3^\nu\g_{A_\alpha A_\mu A_\nu}(q_1,q_2,q_3) & = & 
\left[i\Delta^{(-1)\,\rho}_{\alpha}(q_1)+q_{1}^{\rho}q_{1\alpha}\right]
\left[q_3^2D(q_3)\right] H_{\rho \mu}(q_1,q_2) \nonumber \\
& - & \left[i\Delta^{(-1)\,\rho}_{\mu}(q_2)+q_2^\rho q_{2\mu}\right]
\left[q_3^2D(q_3)\right] H_{\rho\alpha}(q_2,q_1),
\label{sti3gv}
\eea
which, at tree-level, assumes the simple form 
\be
q_3^\nu\g^{(0)}_{A_\alpha A_\mu
A_\nu}(q_1,q_2,q_3)=g\left[g_{\alpha\mu}q_2^2-q_{2\alpha}q_{2\mu}
\right]-g\left[g_{\alpha\mu}q_1^2-q_{1\alpha}q_{1\mu}
\right].
\label{sti3gv0}
\ee

In the BV formalism, the corresponding STI satisfied by the 
three-gluon vertex  $\g_{A^a_\alpha A^b_\mu A^c_\nu}(q_1,q_2)$ may be obtained 
by considering the following functional differentiation of the STI functional
of Eq.\r{mequ}:
\be
\left.\frac{\delta^3{\cal S}\left(\g\right)}{\delta c^c(q_3)\delta
A_\mu^b(q_2)\delta A_\alpha^a(q_1)}\right|_{\Phi=0}=0 \qquad q_1+q_2+q_3=0, 
\ee
which in turn gives the STI
\bea
& & \g_{c^c{A^{*,d}_\nu}}(-q_3)\g_{{A^{d,\nu}}{A^b_\mu}
{A^a_\alpha}}(q_2,q_1)+
\g_{c^c{A^{*,d}_\nu}{A^a_\alpha}}(q_2,q_1)
\g_{{A^{d,\nu}}{A^b_\mu}}(q_2) \nonumber \\
&+ & 
\g_{c^c{A^{*,d}_\nu}{A^b_\mu}}(q_1,q_2)\g_{{A^{d,\nu}}{A^a_\alpha}}(q_1)=0.
\label{stitgv}
\eea

We can then establish the following identifications
[recall Eq.\r{uns}] 
\be
\g_{{A^a_\mu}{A^b_\nu}}(q)=\delta^{ab}
\left[iq_\mu q_\nu-\Delta^{(-1)}_{\mu\nu}(q)\right] \qquad
\Longrightarrow \qquad
\left\{\begin{array}{l}
\g_{A_\mu A_\nu}^{(0)}(q)=
-iq^2P_{\mu\nu}(q), \\ 
\g_{A_\mu A_\nu}^{(n)}(q)=\Pi^{(n)}_{\mu\nu}(q^2).
\label{idena}
\end{array}\right.
\ee
Moreover we can factor out the Lorentz and group structure of the
two-point function $\g_{c^aA_\mu^{*,b}}(p)$ appearing in Eq.\r{stitgv} to get
\be
\g_{c^aA^{*,b}_\mu}(p)=-i\delta^{ab}p_\mu\g_{cA^*}(p) \qquad
\Longrightarrow \qquad \g_{cA^*}(p)=i\frac{p^\mu}{p^2}\g_{cA^*_\mu}(p).
\label{Idb}
\ee
It is then easy to show that the scalar quantity $\g_{cA}(p)$ is
related to the ghost propagator by the following equation
\be
\g_{cA^*}(p)=-\left[p^2 D(p)\right]^{(-1)}.
\ee
Using the Feynman 
rules derived from ${\cal L}_{\rm BRST}$ (see
Fig.\ref{fig0}) to factor out the color structure function, we find
\be
q_3^\nu\g_{A_\alpha A_\nu A_\mu}(q_1,q_2)\g_{cA^*}(q_3)=
i\g_{cA^{*}_{\rho} A_\alpha}(q_2,q_1)\g_{A^\rho A_\mu}(q_2)-
i\g_{cA^{*}_{\rho} A_\mu}(q_1,q_2)\g_{A^\rho A_\alpha}(q_1),
\ee
which implies the STI of Eq.(\ref{sti3gv}), after the 
following identification 
\be
\g_{cA^*_\alpha A_\beta}(q_1,q_2)\equiv
H_{\alpha\beta}(q_1,q_2).
\label{HvsG}
\ee 
This last relation will be helpful in making contact between 
the quantities appearing in the conventional STI  
formulated in the standard covariant gauges 
(which, as such, have no a-priori
knowledge of the BV formalism) and quantities 
appearing in the BQI derived within the BV 
scheme. 

\subsection{\label{subBQI} Background--Quantum Identities}

The BQIs were first presented in \cite{Gambino:1999ai,Grassi:2001zz}
in the context of the Standard Model; 
they may be derived by appropriate functional differentiation
of the BFM STI functional of~Eq.\r{mequbfm}~\cite{Ack} .

\subsubsection{Gluon two-point function}

Consider the following functional differentiation of the STI functional
Eq.\r{mequ} 
\bea
\left.\frac{\delta^2{\cal S}\left(\g\right)}{\delta
\Omega_\alpha^a(p_1)\delta\widehat{A}_\beta^b(q)}\right|_{\Phi=0}=0 
&\qquad& q+p_1=0, \nonumber\\
\left.\frac{\delta^2{\cal S}\left(\g\right)}{\delta
\Omega_\alpha^a(p_1)\delta A_\beta^b(q)}\right|_{\Phi=0}=0 &\qquad& q+p_1=0, 
\eea
which will give the BQIs
\bea
\g_{\widehat{A}_\alpha^a\widehat{A}_\beta^b}(q)&=&\left[g_{\alpha\rho}
\delta^{ad}+ 
\g_{\Omega_\alpha^a A^{*,d}_\rho}(-p_1)\right]\g_{A^{d,\rho} 
\widehat A^b_\beta}(q),  \label{twoBQI1}\\
\g_{\widehat{A}_\alpha^a A_\beta^b}(q)&=&\left[g_{\alpha\rho}\delta^{ad}+
\g_{\Omega_\alpha^a A^{*,d}_\rho}(-p_1)\right]\g_{A^{d,\rho} A^b_\beta}(q).
\label{twoBQI2}
\eea

We can now combine Eqs.\r{twoBQI1} and \r{twoBQI2} in such a way that
the two-point function mixing background and quantum fields drops out
\cite{Grassi:2001zz}. Factoring out the gauge group invariant tensor  
$\delta^{ab}$ and the Lorentz transverse projector
$P_{\alpha\beta}(q)$, we then arrive to the equation
\be
\g_{\widehat{A}\widehat{A}}(q)=\left[1+
\g_{\Omega A^*}(q)\right]^2\g_{AA}(q).
\label{bvsqtwopf}
\ee

The quantity $\g_{\Omega A^*}(q)$ may be constructed 
order-by order using the
Feynman rules derived from ${\cal L}_{\rm BRST}$, listed in Fig.\ref{fig0}.

\begin{figure}[!t]

\bce
\bpi(0,45)(130,-20)

\Line(20,0.75)(40,0.75)
\Line(20,-0.75)(40,-0.75)
\Photon(60,20)(40,0){1.5}{6}
\DashArrowLine(60,-20)(40,0){1}

\Text(20,-8)[l]{\footnotesize{$A^{*,a}_\alpha$}}
\Text(62.5,20.5)[l]{\footnotesize{$A^b_\beta$}}
\Text(62.5,-19.5)[l]{\footnotesize{$c^c$}}
\Text(65,0)[l]{$=-ig f^{abc}g_{\alpha\beta}$}

\Line(170,0.75)(190,0.75)
\Line(170,-0.75)(190,-0.75)
\Photon(190,0)(210,20){1.5}{6}
\DashArrowLine(190,0)(210,-20){1}

\Text(170,-8)[l]{\footnotesize{$\Omega^a_\alpha$}}
\Text(212.5,20.5)[l]{\footnotesize{$A^b_\beta$}}
\Text(212.5,-19.5)[l]{\footnotesize{$\bar c^c$}}
\Text(215,0)[l]{$=-ig f^{abc}g_{\alpha\beta}$}

\epi
\ece

\caption{\label{fig0} Feynman rules from which the two- and
three-point functions
$\g^{(n)}_{\Omega_\alpha A^*_\beta}(q_1)$ and $\g^{(n)}_{cA^*_\alpha
A_\beta}(q_1,q_2)$ can be built up.}
\end{figure}

The crucial point, which allows the exploitation of the 
BQIs derived above, is the observation that $\g_{\Omega A^*}(q)$
may be written in terms of the amplitudes  $D$, $\Delta$, and
most importantly $H$, which appear in the STI for the three-gluon 
vertex, and 
are defined in the context of the conventional formalism, {\it i.e.}, 
have no a-priori knowledge of anti-fields, or of the Feynman rules
stemming from ${\cal L}_{\rm BRST}$.

In particular we have that
the following Schwinger-Dyson equation
holds (perturbatively)
\bea
i\g^{(n)}_{\Omega_\alpha A^*_\mu}(q)=C_A
\int\!\frac{d^4 k}{(2\pi)^4}
\g^{(0)}_{c\Omega_\alpha A_\mu}(q,-k-q)
D^{(n_1)}(k)\Delta^{(n_2)\,\mu\nu}(k+q)
\g^{(n_3)}_{cA^*_\beta A_\nu}(-q,k+q),\nonumber \\
\label{gpert1}
\eea
or diagrammatically
\be
\bpi(0,70)(75,-35)

\Line(-10,0.75)(15,0.75)
\Line(-10,-0.75)(15,-0.75)
\PhotonArc(40,0)(25,0,180){-1.5}{8.5}
\DashCArc(40,0)(25,180,360){1}
\Line(65,0.75)(105,0.75)
\Line(65,-0.75)(105,-0.75)
\GCirc(65,0){15}{0.8}
\GCirc(40,23.5){12}{0.8}
\GCirc(40,-25){12}{0.8}
\DashArrowLine(18.2,-12.3)(19.2,-13.7){1}
\DashArrowLine(56.2,-18.8)(57.2,-17.8){1}

\Text(66,0)[c]{$\scriptstyle{\g^{(n_3)}_{cA^*_\beta A_\nu}}$}
\Text(40,-25)[c]{$\scriptstyle{D^{(n_1)}}$}
\Text(40,23.5)[c]{$\scriptstyle{\Delta^{(n_2)}_{\mu\nu}}$}
\Text(-10,7)[l]{\footnotesize{${\Omega_\alpha}$}}
\Text(105,7)[r]{\footnotesize{${A^*_\beta}$}}
\Text(-20,-1)[r]{$i\g^{(n)}_{\Omega_\alpha A^*_\beta}(q)\,=$}
\Text(150,0)[l]{{$n=n_1+n_2+n_3+1$}}
\epi
\ee

Clearly, from the basic Feynman rules of 
${\cal L}_{\rm BRST}$, we have that 
$\g_{c A^*_\alpha A_\beta}^{(0)}(q_1,q_2) = 
\g_{c \Omega_\alpha A_\beta}^{(0)}(q_1,q_2)$. 
Then using Eq.\r{HvsG} we find
\be
i\g^{(n)}_{\Omega_\alpha A^*_\beta}(q)=C_A 
\int\!\frac{d^4 k}{(2\pi)^4}
H^{(0)}_{\alpha\mu}(q,-k-q)
D^{(n_1)}(k)\Delta^{(n_2)\,\mu\nu}(k+q)H^{(n_3)}_{\beta\nu}(-q,k+q),
\label{gpert2}
\ee
or, diagrammatically,
\be
\bpi(0,70)(75,-35)

\PhotonArc(40,0)(25,0,180){-1.5}{8.5}
\DashCArc(40,0)(25,180,360){1}
\Photon(65,0)(105,0){1.5}{5}
\GCirc(65,0){15}{0.8}
\GCirc(40,23.5){12}{0.8}
\GCirc(40,-25){12}{0.8}
\DashArrowLine(18.2,-12.3)(19.2,-13.7){1}
\DashArrowLine(56.2,-18.8)(57.2,-17.8){1}
\Vertex(15,0){1.8}
\Text(20,0)[l]{$\scriptstyle{\alpha}$} 

\Text(66,0)[c]{$\scriptstyle{H^{(n_3)}_{\beta\nu}}$}
\Text(40,-25)[c]{$\scriptstyle{D^{(n_1)}}$}
\Text(40,23.5)[c]{$\scriptstyle{\Delta^{(n_2)}_{\mu\nu}}$}
\Text(0,-1)[r]{$i\g^{(n)}_{\Omega_\alpha A^*_\beta}(q)\,=$}
\Text(150,0)[l]{{$n=n_1+n_2+n_3+1$}}
\epi
\ee

Evidently this last equation expresses 
$\g^{(n)}_{\Omega_\alpha A^*_\beta}(q)$, a quantity definable in the
BV framework, entirely in terms of quantities definable in the
conventional formalism.  
Using the diagrammatic definition of $H_{\alpha\mu}(q_1,q_2)$ shown
in Eq.\r{pertexp}, we may express diagrammatically
this last Schwinger-Dyson equation 
in terms of the four-particle kernel 
${\cal K}_{\nu\rho}$
as follows, 
\be
\bpi(0,70)(75,-25)

\Line(-10,0.75)(15,0.75)
\Line(-10,-0.75)(15,-0.75)
\PhotonArc(40,0)(25,0,180){-1.5}{8.5}
\DashCArc(40,0)(25,180,360){1}
\PhotonArc(90,0)(25,0,180){-1.5}{8.5}
\DashCArc(90,0)(25,180,360){1}
\Line(115,0.75)(140,0.75)
\Line(115,-0.75)(140,-0.75)
\DashArrowLine(18.2,-12.3)(19.2,-13.7){1}
\DashArrowLine(56.2,-18.8)(57.2,-17.8){1}
\DashArrowLine(110.8,-13.8)(111.7,-12.3){1}
\DashArrowLine(72.6,-17.8)(73.6,-18.8){1}

\GCirc(90,23.5){12}{0.8}
\GCirc(90,-25){12}{0.8}
\GCirc(65,0){12}{0.8}
\GCirc(40,23.5){12}{0.8}
\GCirc(40,-25){12}{0.8}

\Text(66,0)[c]{$\scriptstyle{{\cal K}^{(\tilde n_3)}_{\nu\rho}}$}
\Text(40,-25)[c]{$\scriptstyle{D^{(n_1)}}$}
\Text(40,23.5)[c]{$\scriptstyle{\Delta^{(n_2)}_{\mu\nu}}$}
\Text(90,-25)[c]{$\scriptstyle{D^{(\tilde n_1)}}$}
\Text(90,23.5)[c]{$\scriptstyle{\Delta^{(\tilde n_2)}_{\rho\sigma}}$}

\Text(-10,7)[l]{\footnotesize{${\Omega_\alpha}$}}
\Text(140,7)[r]{\footnotesize{${A^*_\beta}$}}
\Text(-20,-1)[r]{$i\g^{(n)}_{\Omega_\alpha A^*_\beta}(q)\,=$}

\epi
\ee

\subsubsection{Gluon--quark--anti-quark three-point function}

For the annihilation channel (one can study equally well the elastic channel)
we consider the functional differentiation
\be
\left.\frac{\delta^3{\cal S}\left(\g\right)}{\delta \Omega^a_\alpha(q)\delta
\bar\psi(Q')\delta \psi(Q)}\right|_{\Phi=0}=0 \qquad Q'+Q+q=0, 
\ee
which provides us the BQI
\bea
\g_{\widehat A^a_\alpha\bar\psi\psi}(Q',Q)&=&\left[g_{\alpha\rho}\delta^{ad}+
\g_{\Omega^a_\alpha A^{*,d}_\rho}(-q)\right]
\g_{A^{d,\rho}\bar\psi\psi}(Q',Q)+\g_{\bar\psi\psi}(-Q')
\g_{\Omega_\alpha^a \bar\psi^*\psi}(Q',Q) \nonumber \\
&-&\g_{\Omega_\alpha^a \psi^*\bar\psi}(Q,Q')
\g_{\bar\psi\psi}(Q). 
\label{thpfbqi}
\eea
Since we will always deal with on-shell external fermions, we can sandwich the
above equation between on-shell Dirac spinors; in this way, using the
Dirac equations of motion 
$\g_{\bar\psi\psi}(Q)u(Q)=0$ and $\bar v(Q')\g_{\bar\psi\psi}(-Q')=0$ 
when $\Qsm=\Qpsm=m$, we can get rid of the second and third term appearing in 
Eq.\r{thpfbqi}, to write the on-shell BQI
\be 
\g_{\widehat A^a_\alpha\bar\psi\psi}(Q',Q)
=\left[g_{\alpha\rho}\delta^{ad}+\g_{\Omega^a_\alpha A^{*,d}_\rho}(q)\right]
\g_{A^{d,\rho}\bar\psi\psi}(Q',Q).
\label{bvsqthreepf}
\ee

The reader should appreciate the fact that the BQIs alone, interesting
as they may be in their  own right, would be of limited usefulness for
our purposes, if  it were not for the  complementary identification of
the  corresponding  pieces  appearing  in  the  STI,  as  captured  in
Eq.(\ref{gpert1}).
Notice in particular that the BQIs
by themselves only amount to the statement that  
$\Gamma^{(n_1)}\Delta^{(n_2)}\Gamma^{(n_3)} = 
\widehat{\Gamma}^{(n_1)}\widehat{\Delta}^{(n_2)} \widehat{\Gamma}^{(n_3)}$, 
which is automatically true, since
the box-diagrams are identical in both schemes, 
and so is the entire $S$-matrix.

\subsubsection{Gluon three-point function}

Here we derive the BQI relating the gluon
three-point function 
$\g_{\widehat{A}_\alpha^a A_\beta^b A_\gamma^c}(p_1,p_2)$, {\it i.e.}, 
with one background and two quantum gluons to the normal 
gluon three-point function 
$\g_{A_\alpha^a A_\beta^b A_\gamma^c}(p_1,p_2)$
{\it i.e.}, with three quantum gluons.

We start by considering the following functional differentiation
\be
\left.\frac{\delta^3{\cal S}\left(\g\right)}{\delta
\Omega_\alpha^a(q)\delta A_\beta^b(p_1)\delta A_\gamma^c(p_2)}
\right|_{\Phi=0}=0 \qquad q+p_1+p_2=0, 
\ee
which will provide us the BQI
\bea
\g_{\widehat{A}_\alpha^a A_\beta^b A_\gamma^c}(p_1,p_2)&=&
\left[g_{\alpha\rho}\delta^{ad}+
\g_{\Omega_\alpha^a A^{*,d}_\rho}(q)\right]\g_{A^{d,\rho} A^b_\beta
A_\gamma^c}(p_1,p_2) \nonumber \\
&+&\g_{\Omega_\alpha^aA^{*,d}_\rho A^c_\gamma}(p_1,p_2)
\g_{A^{d,\rho}A^b_\beta}(p_1)+
\g_{\Omega_\alpha^aA^{*,d}_\rho A^b_\beta}(p_2,p_1)
\g_{A^{d,\rho}A^c_\gamma}(p_2). \nonumber \\
\label{tgBQIOS}
\eea
Next, we consider the case in which the external gluons $A^b_\beta(p_1)$ and 
$A^c_\gamma(p_2)$ are ``on-shell'' physical states, {\it i.e.}, with
$p_1^2=p_2^2=0$  and $p_1^\beta\epsilon_\beta(p_1)
=p_2^\gamma\epsilon_\gamma(p_2)=0$.
Then, since the gluon propagator is transverse, 
we find the on-shell BQI
\be
\g_{\widehat{A}_\alpha^a A_\beta^b A_\gamma^c}(p_1,p_2)=
\left[g_{\alpha\rho}\delta^{ad}+
\g_{\Omega_\alpha^a A^{*,d}_\rho}(q)\right]\g_{A^{d,\rho} A^b_\beta
A_\gamma^c}(p_1,p_2).
\label{tgBQI}
\ee

Notice that in the above BQI
the (unphysical) Green's function that provides the organization 
of the Feynman diagrams for converting the three-point function 
$\g_{\widehat{A}_\alpha^a A_\beta^b A_\gamma^c}(p_1,p_2)$ into the
corresponding  
quantum one  $\g_{A^a_\alpha A^b_\beta A_\gamma^c}(p_1,p_2)$, is the
same that appears 
in the two-point BQI of Eqs.\r{bvsqtwopf} and \r{bvsqthreepf}.

\subsubsection{Gluon four-point function}

Finally, we derive the BQI relating the gluon
four-point function 
$\g_{\widehat{A}_\alpha^a A_\beta^b A_\gamma^c A_\delta^d}(p_1,p_2,p_3)$, 
{\it i.e.}, 
with one background and three quantum gluons to the normal gluon 
four-point function 
$\g_{A_\alpha^a A_\beta^b A_\gamma^c A^d_\delta}(p_1,p_2,p_3)$,
{\it i.e.}, with four quantum gluons.

For doing this we consider the functional differentiation
\be
\left.\frac{\delta^4{\cal S}\left(\g\right)}{\delta
\Omega_\alpha^a(q)\delta A_\beta^b(p_1)\delta A_\gamma^c(p_2)
\delta A_\delta^d(p_3)}
\right|_{\Phi=0}=0 \qquad q+p_1+p_2+p_3=0, 
\ee
which will give us the BQI
\bea
\g_{\widehat{A}_\alpha^a A_\beta^b A_\gamma^c A_\delta^d}(p_1,p_2,p_3)&=&
\left[g_{\alpha\rho}\delta^{ae}+
\g_{\Omega_\alpha^a A^{*,e}_\rho}(q)\right]\g_{A^{e,\rho} A^b_\beta
A_\gamma^c A_\delta^d}(p_1,p_2,p_3) \nonumber \\
&+ &
\g_{\Omega_\alpha^a A^{*,e}_\rho A^b_\beta}(q,p_1)
\g_{A^{e,\rho}A^c_\gamma A^d_\delta}(q+p_1,p_2)\nonumber \\
&+ &
\g_{\Omega_\alpha^a A^{*,e}_\rho A^c_\gamma}(q,p_2)
\g_{A^{e,\rho}A^d_\delta A^b_\beta}(q+p_2,p_3) \nonumber \\
&+ & 
\g_{\Omega_\alpha^a A^{*,e}_\rho A^d_\delta}(q,p_3)
\g_{A^{d,\rho}A^b_\beta A^c_\gamma}(q+p_3,p_1)\nonumber \\
&+ & 
\g_{\Omega^a_\alpha A^b_\beta A^c_\gamma A^{*,e}_\rho}(q,p_1,p_2)
\g_{A^{e,\rho}A^d_\delta}(p_3)\nonumber \\
&+ & 
\g_{\Omega^a_\alpha A^d_\delta A^b_\beta A^{*,e}_\rho}(q,p_3,p_1)
\g_{A^{e,\rho}A^c_\gamma}(p_2)\nonumber \\
&+ & 
\g_{\Omega^a_\alpha A^c_\gamma A^d_\delta A^{*,e}_\rho}(q,p_2,p_3)
\g_{A^{e,\rho}A^b_\beta}(p_1). 
\eea
If, as before, the external gluons  $A^b_\beta(p_1)$, 
$A^c_\gamma(p_2)$ and $A^d_\delta(p_3)$ are considered 
as on-shell physical states, we can 
get rid of the last three terms, obtaining the on-shell BQI
\bea
\g_{\widehat{A}_\alpha^a A_\beta^b A_\gamma^c A_\delta^d}(p_1,p_2,p_3) &=&
\left[g_{\alpha\rho}\delta^{ae}+
\g_{\Omega_\alpha^a A^{*,e}_\rho}(q)\right]\g_{A^{e,\rho} A^b_\beta
A_\gamma^c A_\delta^d}(p_1,p_2,p_3) \nonumber \\
& + & 
\g_{\Omega_\alpha^a A^{*,e}_\rho A^b_\beta}(q,p_1)
\g_{A^{e,\rho}A^c_\gamma A^d_\delta}(q+p_1,p_2)\nonumber \\
& + &
\g_{\Omega_\alpha^a A^{*,e}_\rho A^c_\gamma}(q,p_2)
\g_{A^{e,\rho}A^d_\delta A^b_\beta}(q+p_2,p_3) \nonumber \\
& + & 
\g_{\Omega_\alpha^a A^{*,e}_\rho A^d_\delta}(q,p_3)
\g_{A^{e,\rho}A^b_\beta A^c_\gamma}(q+p_3,p_1).
\label{fgBQI}
\eea

Again we find the same unphysical Green's function emerging, plus three terms 
that where not present (due to the on-shell condition) 
in Eq.\r{tgBQI}.

\section{\label{sec:three} A fresh look at the $S$-matrix PT}

In the next two sections we will review the $S$-matrix PT
in an attempt to accomplish two main objectives. 
First, we will furnish a discussion of the method, 
which incorporates into a coherent framework the
various conceptual and technical development which 
have taken place in the last years. Second, 
we use it as an opportunity to   
familiarize ourselves with the
BV formalism, and in particular the BQIs, 
in a well-understood context. Thus,  
after outlining the general PT framework, we  
will re-express the one-loop
$S$-matrix PT results in terms of the BV building blocks.
The more technical case of the two-loop PT, together with 
the corresponding BV ingredients, will be revisited 
in the next section. 

\subsection{General framework}

A general $S$-matrix element of a $2\to 2$ process
can be written following 
the standard Feynman rules as 
\be
T(s,t,m_i)\ =\ T_1(s,\xi)\ +\ T_2(s,m_i,\xi)\ +\
T_3(s,t,m_i,\xi),
\label{Arx}
\ee
Evidently the Feynman diagrams impose 
a decomposition of  $T(s,t,m_i)$ into three distinct sub-amplitudes 
$T_1$, $T_2$, and $T_3$, with a very characteristic kinematic
structure, {\it i.e.}, a very particular dependence on the the 
Mandelstam kinematic
variables and the masses. Thus, $T_1$ 
is the conventional self-energy contribution, which
only depends on
the momentum transfer $s$, $T_2$ corresponds to vertex diagrams
which in general depend also on the masses of the external particles, 
whereas $T_3$ is a box-contribution, having in addition a non-trivial
dependence on the Mandelstam variable $t$. However, all these 
sub-amplitudes, in addition to their dependence of the physical 
kinematic variables, also display a non-trivial dependence on
the unphysical gauge fixing parameter parameter $\xi$. 
Of course 
we know that the BRST symmetry guarantees that 
the total $T(s,t,m_i)$ is independent of 
$\xi$, {\it i.e.}, $d T/d \xi =0$; thus, in general, a  
set of delicate gauge-cancellations will take place. 
The PT framework provides a very particular realization of this
cancellations.
Specifically, the transition
amplitude $T(s,t,m_i)$ of a $2\to 2$ process,
can be decomposed as
\cite{Cornwall:1982zr,Cornwall:1989gv,Papavassiliou:1990zd} 
\be
T(s,t,m_i)\ =\ \widehat{T}_1(s)\ +\ \widehat{T}_2(s,m_i)\ +\
\widehat{T}_3(s,t,m_i),
\label{TPT}
\ee
in terms of three individually gauge-invariant
quantities:
a propagator-like part ($\widehat{T}_1$), a vertex-like piece
($\widehat{T}_2$),
and a part containing box graphs ($\widehat{T}_3$). The important observation
is that vertex and box graphs contain in general
pieces, which are kinematically akin to self-energy graphs
of the transition amplitude.
The PT is a systematic way of extracting such pieces and
appending them to the conventional self-energy graphs.
In the same way, effective gauge invariant
vertices may be constructed, if
after subtracting from the conventional vertices the
propagator-like pinch parts we add the vertex-like pieces, if any, 
coming from boxes. 
The remaining purely box-like contributions are then
also gauge invariant. 
In what follows we will consider for concreteness 
the $S$-matrix element for the  
quark ($q$)-antiquark ($\bar{q}$) elastic scattering process 
$q(P)\bar{q}(P')\to q(Q)\bar{q}(Q')$ in QCD;
we set $q= P'-P= Q'-Q$, with $s=q^2$ is  
the square of the momentum transfer. One could equally well
study the annihilation channel, in which case $s$ would be
the center-of-mass energy. 

In order to identify the pieces which are
to be reassigned, all one has to do is to resort to the fundamental
WIs of the theory. In particular the longitudinal
momenta $k_{\mu}$ appearing inside Feynman diagrams eventually
reach the elementary gluon-quark vertex involving
one ``on-shell'' quark carrying momentum $Q$
and one off-shell quark, carrying momentum $k+Q$,
and trigger the WI 
\bea
k_{\mu} [\bar{u}(\Qsm)\, \gamma^{\mu}\, S(\ksm + \Qsm)] &=& 
\bar{u}(\Qsm)\, \ksm \, S(\ksm + \Qsm) \nonumber\\
 &=& \bar{u}(\Qsm)\,[ (\ksm + \Qsm +m ) - (\Qsm +m)]\,S(\ksm + \Qsm)
\nonumber\\
&=& \bar{u}(\Qsm)[\,S^{-1}(\ksm + \Qsm)- \,S^{-1}(\Qsm)]\, S(\ksm + \Qsm)
\label{FWI}
\eea
The first term in the square bracket will remove (pinch out) the
internal quark propagator, giving rise to a self-energy-like
contributions, while the second term will die on-shell, by virtue
of the Dirac equation of motion; the on-shell condition used 
at this point is characteristic of the $S$-matrix PT 
\cite{Cornwall:1982zr,Cornwall:1989gv,Papavassiliou:1990zd}.  

An important step in the PT procedure is clearly the identification 
of all {\it longitudinal} momenta
involved, {\it i.e.}, the momenta which can trigger 
the elementary WI of Eq.(\ref{PTWI}).
There are two sources of such momenta: ({\it i}) The tree-level expressions
for the gauge boson propagators appearing inside Feynman diagrams
and ({\it ii}) the tri-linear gauge boson vertices. Regarding the former 
contributions, the tree-level gluon propagator reads
\be
\Delta_{\mu\nu}(q) = \frac{-i}{q^2}\left[g_{\mu\nu} -
(1-\xi)\frac{q_{\mu}q_{\nu}}{q^2}\right], 
\ee
and the longitudinal momenta are simply those multiplying 
$(1-\xi)$. 
It is a straightforward but tedious 
exercise to convince one-self that inside an 
$S$-matrix element all terms proportional
to $(1-\xi)^n$, with $n \geq 1$, cancel against each-other
in a very special way. 
In particular,
all relevant cancellations proceed without  need of carrying out
integrations over the virtual loop momenta, thus maintaining the kinematic
identity of the various  Green's functions intact,
a point of crucial importance within the PT philosophy.  
As has been shown by explicit calculations 
(see for example \cite{Binosi:2001hy}), 
this is indeed 
the case at one- and two-loops.
The key
observation
is  that all contributions
originating from the longitudinal parts of gauge boson propagators, by
virtue of the WIs they trigger,  give rise to {\it unphysical} effective
vertices,  {\it i.e.}, vertices which do not exist in  the   original 
Lagrangian. All  such  vertices   cancel  {\it diagrammatically} inside
ostensibly gauge-invariant  quantities, such as current correlation functions
or $S$-matrix elements. 
It is important to
emphasize  that exactly the same result is obtained 
even in  the context  of
the non-covariant  axial gauges  (see for example
\cite{Dokshitzer:1980hw,Andrasi:1981rr,Capper:1982rd}), 
where the  Feynman gauge cannot be  reached a priori  by simply fixing
appropriately the value of the  gauge-fixing parameter.  Thus, even if
one  uses a bare  gluon propagator  of the  general axial  gauge form,
after the  aforementioned cancellations  have taken place  one arrives
effectively to the answer written in the covariant Feynman gauge. 
Thus,  one can
begin the analysis without loss of generality
by  choosing   the Feynman gauge when  
writing  down  the   Feynman  diagrams contributing  to the  $S$-matrix.

The identification and role of the longitudinal momenta stemming
from the three-gluon vertex is slightly more subtle. 
The fundamental tree-level three-gluon vertex 
$\g_{A^a_\alpha A^b_\mu A^c_\nu}^{(0)}(q,p_1,p_2)$ is given by the
following manifestly Bose-symmetric expression
(all momenta are incoming, {\it i.e.}, $q+p_1+p_2 = 0$)
\bea
& &\g_{A^a_\alpha A^b_\mu A^c_\nu}^{(0)}(q,p_1,p_2)=g f^{abc}
\Gamma_{\alpha \mu \nu}^{(0)}(q,p_1,p_2), \nonumber \\
& & \Gamma_{\alpha \mu \nu}^{(0)}(q,p_1,p_2)= 
(q-p_1)_{\nu}g_{\alpha\mu} + (p_1-p_2)_{\alpha}g_{\mu\nu}
 + (p_2-q)_{\mu}g_{\alpha\nu}.
\eea
The Lorentz structure $\Gamma_{\alpha \mu \nu}^{(0)}(q,p_1,p_2)$
may be split into two parts 
\cite{'tHooft:1971fh, Cornwall:1977ii}
\be
\Gamma_{\alpha \mu \nu}^{(0)}(q,p_1,p_2) 
= 
\Gamma_{\alpha \mu \nu}^{{\rm F}}(q,p_1,p_2) + 
\Gamma_{\alpha \mu \nu}^{{\rm P}}(q,p_1,p_2),
\label{decomp}
\ee
with 
\bea
\Gamma_{\alpha \mu \nu}^{{\rm F}}(q,p_1,p_2) &=& 
(p_1-p_2)_{\alpha} g_{\mu\nu} + 2q_{\nu}g_{\alpha\mu} 
- 2q_{\mu}g_{\alpha\nu} \, , \nonumber\\
\Gamma_{\alpha \mu \nu}^{{\rm P}}(q,p_1,p_2) &=&
 p_{2\nu} g_{\alpha\mu} - p_{1\mu}g_{\alpha\nu}.  
\label{GFGP}
\eea
The vertex $\Gamma_{\alpha \mu \nu}^{{\rm F}}(q,p_1,p_2)$ 
is Bose-symmetric only with respect to the
$\mu$ and $\nu$ legs, and 
coincides with the BFM Feynman gauge bare vertex involving one
background gluon (carrying four-momentum
$q$) and two quantum gluons (carrying four-momenta $p_1$ and $p_2$). 
Evidently the above decomposition assigns a special r\^ole 
to the $q$-leg,
and allows $\Gamma_{\alpha \mu \nu}^{{\rm F}}(q,p_1,p_2)$ to satisfy the WI
\be 
q^{\alpha} \Gamma_{\alpha \mu \nu}^{{\rm F}}(q,p_1,p_2) = 
(p_2^2 - p_1^2)g_{\mu\nu},
\label{WI2B}
\ee
where the right-hand side (RHS) is the difference of two-inverse
propagators in the renormalizable Feynman gauge. 
The term $\Gamma_{\alpha \mu \nu}^{{\rm P}}(q,p_1,p_2)$
contains the pinching momenta; they will eventually trigger the
elementary WI, which will eliminate the internal quark propagator,
resulting in an effectively propagator-like contribution.
Notice that in the light of the BV formalism, the PT splitting given in 
Eq.(\ref{GFGP}) may be cast in the alternative, perhaps 
more suggestive form
\be
\g_{A^a_\alpha A^b_\mu A^c_\nu}^{(0)}(q,p_1,p_2) =
\g_{\widehat{A}^a_\alpha A^b_\mu A^c_\nu}^{(0)}(q,p_1,p_2)
+  i \left[p_{2\nu}\g^{(0)}_{cA^*_\alpha A_\mu}  
- p_{1\mu}\g^{(0)}_{cA^*_\alpha A_\nu}\right].
\ee

\begin{figure}[!t]
\bce
\begin{picture}(0,95)(155,-35)

\Text(5,0)[r]{$\g_{A_\alpha\bar\psi\psi}(Q',Q)\,=$}
\Text(97.5,0)[c]{\footnotesize{$+$}}
\Text(197.5,0)[c]{\footnotesize{$+$}}
\Text(297.5,0)[c]{\footnotesize{$+$}}

\Text(15,-50)[l]{\footnotesize{(a)}}
\Text(115,-50)[l]{\footnotesize{(b)}}
\Text(215,-50)[l]{\footnotesize{(c)}}
\Text(315,-50)[l]{\footnotesize{(d)}}

%A

\Photon(15,0)(33.5,0){1.5}{3}
\PhotonArc(50,0)(15,-8,352){1.5}{14}
\ArrowLine(85,20)(65,0)
\ArrowLine(65,0)(85,-20)
\GCirc(65,0){8}{0.8}
\GCirc(50,13.5){5}{0.8}
\Text(50,13.5)[c]{$\scriptstyle{\Delta}$}
\GCirc(50,-13.5){5}{0.8}
\Text(50,-13.5)[c]{$\scriptstyle{\Delta}$}
\Text(65,0)[c]{\scriptsize{1PI}}

%B

\Photon(115,25)(133.5,25){1.5}{3}
\DashCArc(150,25)(15,-8,352){1}
\ArrowLine(160.51,35.61)(160.61,35.51)
\ArrowLine(139.29,35.51)(139.39,35.61)
\ArrowLine(139.29,14.49)(139.39,14.39)
\ArrowLine(160.51,14.39)(160.61,14.49)
\ArrowLine(185,45)(165,25)
\ArrowLine(165,25)(185,5)
\GCirc(165,25){8}{0.8}
\GCirc(150,38.5){5}{0.8}
\Text(150.5,38.5)[c]{$\scriptstyle{D}$}
\GCirc(150,11.5){5}{0.8}
\Text(150.5,11.5)[c]{$\scriptstyle{D}$}
\Text(165,25)[c]{\scriptsize{1PI}}

\Photon(115,-25)(133.5,-25){1.5}{3}
\DashCArc(150,-25)(15,-8,352){1}
\ArrowLine(160.61,-14.49)(160.51,-14.39)
\ArrowLine(139.39,-14.39)(139.29,-14.49)
\ArrowLine(139.39,-35.61)(139.29,-35.51)
\ArrowLine(160.61,-35.51)(160.51,-35.61)
\ArrowLine(185,-5)(165,-25)
\ArrowLine(165,-25)(185,-45)
\GCirc(165,-25){8}{0.8}
\GCirc(150,-11.5){5}{0.8}
\Text(150.5,-11.5)[c]{$\scriptstyle{D}$}
\GCirc(150,-38.5){5}{0.8}
\Text(150.5,-38.5)[c]{$\scriptstyle{D}$}
\Text(165,-25)[c]{\scriptsize{1PI}}

%C

\Photon(215,0)(233.5,0){1.5}{3}
\Photon(233.5,0)(260,0){1.5}{3.5}
\PhotonArc(250,0)(15,-8,352){1.5}{12}
\ArrowLine(280,20)(265,5)
\ArrowLine(265,-5)(280,-20)
\GCirc(265,0){8}{0.8}
\Text(265.5,0)[c]{\scriptsize{1PI}}
\GCirc(250,13.5){5}{0.8}
\Text(250.5,13.5)[c]{$\scriptstyle{\Delta}$}
\GCirc(250,-13.5){5}{0.8}
\Text(250.5,-13.5)[c]{$\scriptstyle{\Delta}$}
\GCirc(247.5,0){5}{0.8}
\Text(248,0)[c]{$\scriptstyle{\Delta}$}

%D

\Photon(315,25)(333.5,25){1.5}{3}
\CArc(350,25)(15,-8,352)
\ArrowLine(360.51,35.61)(360.61,35.51)
\ArrowLine(339.29,35.51)(339.39,35.61)
\ArrowLine(339.29,14.49)(339.39,14.39)
\ArrowLine(360.51,14.39)(360.61,14.49)
\ArrowLine(385,45)(365,25)
\ArrowLine(365,25)(385,5)
\GCirc(365,25){8}{0.8}
\GCirc(350,38.5){5}{0.8}
\Text(350.5,38.5)[c]{$\scriptstyle{S}$}
\GCirc(350,11.5){5}{0.8}
\Text(350.5,11.5)[c]{$\scriptstyle{S}$}
\Text(365,25)[c]{\scriptsize{1PI}}

\Photon(315,-25)(333.5,-25){1.5}{3}
\CArc(350,-25)(15,-8,352)
\ArrowLine(360.61,-14.49)(360.51,-14.39)
\ArrowLine(339.39,-14.39)(339.29,-14.49)
\ArrowLine(339.39,-35.61)(339.29,-35.51)
\ArrowLine(360.61,-35.51)(360.51,-35.61)
\ArrowLine(385,-5)(365,-25)
\ArrowLine(365,-25)(385,-45)
\GCirc(365,-25){8}{0.8}
\GCirc(350,-11.5){5}{0.8}
\Text(350.5,-11.5)[c]{$\scriptstyle{S}$}
\GCirc(350,-38.5){5}{0.8}
\Text(350.5,-38.5)[c]{$\scriptstyle{S}$}
\Text(365,-25)[c]{\scriptsize{1PI}}

\end{picture}

\end{center}

\caption{\label{vdeco} The decomposition of the three-point function 
$\g_{A_\alpha\bar\psi\psi}(Q',Q)$ 
in terms of 
diagrams
having an external elementary three-gluon vertex
$\Gamma^{A^2}_{A_\alpha\bar\psi\psi}(Q',Q)$ (a), those where the external
gluon couples directly to ghost fields 
$\Gamma^{\bar c c}_{A_\alpha\bar\psi\psi}(Q',Q)$
(b), and the the rest, which falls into
neither of the previous categories 
$\Gamma^{A^3}_{A_\alpha\bar\psi\psi}(Q',Q)$
and $\Gamma^{\bar q q}_{A_\alpha\bar\psi\psi}(Q',Q)$ 
[(c) and (d) respectively]. $S$ represents the full fermionic propagator.}
\end{figure}
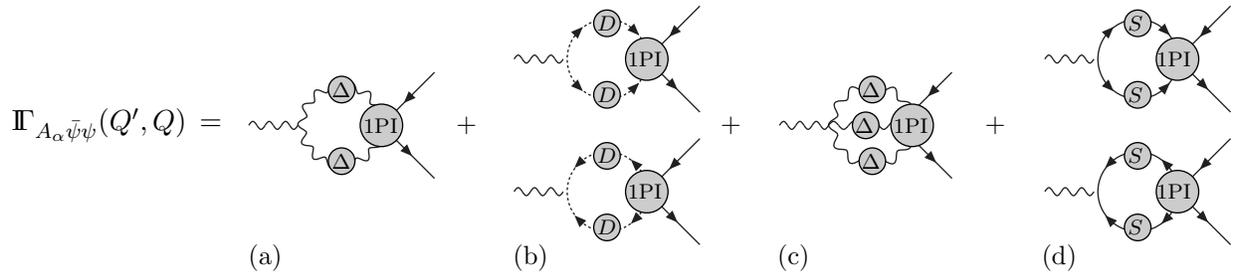

According to the PT \cite{Papavassiliou:1999az,Papavassiliou:1999bb} 
the next steps 
consists of the following: ({\it a}) Classify all
diagrams which contribute to the 
three-point function $\g_{A_\alpha\bar\psi\psi}(Q',Q)$ into
the following categories: ({\it i}) those 
containing an {\it external} three-gluon vertex, 
{\it i.e.}, a three-gluon vertex where the momentum $q$ is incoming,
({\it ii})
those which do not have such an external three-gluon vertex.
This latter set contains graphs where the incoming 
gluon couples to the rest of the diagram 
with any type of interaction vertex other than a three-gluon
vertex. Thus we write (see also Fig.\ref{vdeco})
\be 
\g_{A_\alpha\bar\psi\psi}(Q',Q)=
\Gamma^{A^2}_{A_\alpha\bar\psi\psi}(Q',Q)+
\Gamma^{\bar c c}_{A_\alpha\bar\psi\psi}(Q',Q)+
\Gamma^{A^3}_{A_\alpha\bar\psi\psi}(Q',Q)+
\Gamma^{\bar q q}_{A_\alpha\bar\psi\psi}(Q',Q).
\label{deco1}
\ee
({\it b}) Carry out inside the class ({\it i})
diagrams the vertex decomposition given 
in Eq.(\ref{decomp}). ({\it c}) Track down the terms originating from
$\Gamma_{\alpha \mu \nu}^{{\rm P}}(q,p_1,p_2)$: these terms,
depending on the topological details of the diagram under consideration
will either ({\it i}) trigger directly the WI of Eq.(\ref{FWI}) or
({\it ii})
they will trigger a chain of intermediate tree-level WIs, 
such as Eq.(\ref{STIbc})
whose end result will
be that eventually 
an appropriate longitudinal momentum will be generated, which 
will trigger 
the WI of Eq.(\ref{FWI}).
({\it d}) The propagator-like terms thusly generated are to be alloted 
to the conventional self-energy graphs, and will form part of the
effective PT gluon self-energy at that order; to complete
its construction one needs to supply in addition the left-over pieces
generated when converting a string of 1PI self-energies
into a corresponding PT string. Finally, 
the remaining 
purely vertex-like parts define the effective PT 
gluon-quark-antiquark 
three-point function $\widehat{\g}_{A_\alpha\bar\psi\psi}(Q',Q)$ .  

Before entering into some of the details of the explicit 
one- and two-loop
constructions we would like to comment on an additional subtle point.
One of the main obstacle related to the 
generalization of the PT beyond one-loop 
has been the issue of whether or not a splitting analogous to that
of Eq.(\ref{decomp}) should take place for the
internal three-gluon
vertices, {\it i.e.}, vertices with all three
legs irrigated by virtual momenta, so that $q$ never enters  
{\it alone} into any of the legs.
This issue has been resolved by resorting to the 
special unitarity properties satisfied by the PT Green's functions.
The final answer is 
that no splitting should take place 
for {\it any} of these internal three-gluon
vertices. As we will see in the next section, a new and more direct 
argument corroborates this answer.

\subsection{The one-loop construction}

\begin{figure}[!t]
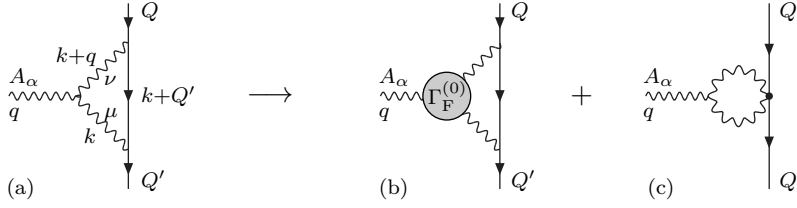

\bce
\bpi(0,50)(170,-20)

\Photon(15,0)(40,0){1.5}{6}
\Photon(60,20)(40,0){-1.5}{6}
\Photon(60,-20)(40,0){1.5}{6}
\ArrowLine(60,35)(60,20)
\ArrowLine(60,20)(60,-20)
\ArrowLine(60,-20)(60,-35)

\Photon(155,0)(180,0){1.5}{6}
\Photon(200,20)(180,0){1.5}{6}
\Photon(200,-20)(180,0){1.5}{6}
\ArrowLine(200,35)(200,20)
\ArrowLine(200,20)(200,-20)
\ArrowLine(200,-20)(200,-35)
\GCirc(180,0){9}{0.8}

\Photon(255,0)(278.5,0){1.5}{6}
\PhotonArc(290,0)(10,-8,352){1.5}{12}
\ArrowLine(301.5,35)(301.5,0)
\ArrowLine(301.5,0)(301.5,-35)
\Vertex(301.5,0){1.4}

\Text(105,0)[l]{$\longrightarrow$}
\Text(181,0)[c]{$\scriptstyle{\Gamma^{(0)}_{\rm F}}$}
\Text(227.5,0)[l]{$+$}
\Text(15,7)[l]{$\scriptstyle{A_\alpha}$}
\Text(15,-7)[l]{$\scriptstyle{q}$}
\Text(51,6)[l]{$\scriptstyle{\nu}$}
\Text(51,-7)[l]{$\scriptstyle{\mu}$}
\Text(48,15)[r]{$\scriptstyle{k+q}$}
\Text(48,-15)[r]{$\scriptstyle{k}$}
\Text(65,0)[l]{$\scriptstyle{k+Q'}$}
\Text(65,32)[l]{$\scriptstyle{Q}$}
\Text(65,-32)[l]{$\scriptstyle{Q'}$}
\Text(155,7)[l]{$\scriptstyle{A_\alpha}$}
\Text(155,-7)[l]{$\scriptstyle{q}$}
\Text(205,32)[l]{$\scriptstyle{Q}$}
\Text(205,-32)[l]{$\scriptstyle{Q'}$}
\Text(255,7)[l]{$\scriptstyle{A_\alpha}$}
\Text(255,-7)[l]{$\scriptstyle{q}$}
\Text(306.5,32)[l]{$\scriptstyle{Q}$}
\Text(306.5,-32)[l]{$\scriptstyle{Q'}$}
\Text(14,-35)[l]{\scriptsize{(a)}}
\Text(155,-35)[l]{\scriptsize{(b)}}
\Text(256.5,-35)[l]{\scriptsize{(c)}}

\epi
\ece

\caption{\label{fdeco} Carrying out the fundamental vertex decomposition inside
the three-point function $\Gamma_{A_\alpha\psi\bar\psi}^{A^2\,(1)}(Q',Q)$ (a)
contributing to $\g_{A_\alpha\psi\bar\psi}^{(1)}(Q',Q)$, 
gives rise to the genuine
vertex $\widehat\Gamma_{A_\alpha\bar\psi\psi}^{A^2\,(1)}(Q',Q)$ (b) and a 
self-energy-like
contribution $\frac12V_{\alpha\rho}^{{\rm P}\,(1)}(q)\gamma^\rho$ (c).}
\end{figure}

Notice that at the one-loop level only the first and last term of 
Eq.\r{deco1}
will be present.
We then implement (see Fig.\ref{fdeco}a) the vertex decomposition of 
Eq.\r{decomp},   with $p_{1\mu}=k_{\mu}$, $p_{2\nu}=-(k+q)_{\nu}$, inside the
$\Gamma^{A^2\,(1)}_{A_\alpha\bar\psi\psi}(Q',Q)$ part of Eq.\r{deco1}. 
The $\Gamma^{{\rm P}}_{\alpha\mu\nu}(q,p_1,p_2)$ term
triggers then the elementary WIs 
\bea
& & \ksm  = (\ksm + \Qpsm-m) - (\Qpsm -m), 
\nonumber\\ 
& & \ksm + \qsm = (\ksm + \Qpsm-m) - (\Qsm -m),
\label{PTWI}
\eea
thus, two self-energy like pieces are generated (Fig.\ref{fdeco}c), 
which are to be alloted
to the conventional self-energy. In particular,
\bea
\Gamma_{A_\alpha\bar\psi\psi}^{A^2\,(1)}(Q',Q) &=& 
\widehat\Gamma_{A_\alpha\bar\psi\psi}^{A^2\,(1)}(Q',Q) 
+\frac{1}{2}V_{\alpha\rho}^{{\rm P}\,(1)}(q) \gamma^{\rho}
-X^{(1)}_{1\,\alpha}(Q',Q)\Sigma^{(0)}(Q') \nonumber \\
&- & \Sigma^{(0)}(Q)X^{(1)}_{2\,\alpha}(Q',Q),
\label{PTact}
\eea
where
\bea
& & \widehat\Gamma_{A_\alpha\bar\psi\psi}^{A^2\,(1)}(Q',Q)
=\int_{L_1} J(q,k) \Gamma_{\alpha\mu\nu}^{{\rm F}}(q,k,-k-q)
\gamma^{\mu} S^{(0)}(Q'+k)\gamma^{\nu},\nonumber\\
& & V^{{\rm P}\,(1)}_{\alpha\rho}(q)=2 g_{\alpha\rho}\,\int_{L_1} J(q,k),
\eea
and 
\bea
\int_{L_1} &\equiv& \mu^{2\varepsilon}\int\!\frac{d^dk}{(2\pi)^d},
\nonumber\\
J(q,k) &\equiv& g^2 C_A [k^2 (k+q)^{2}]^{-1}.
\eea
$C_A$ denotes the Casimir eigenvalue of the adjoint
representation, {\it i.e.}, $C_A = N$ for $SU(N)$. 
Notice that the last two terms appearing in the RHS of 
Eq.\r{PTact} vanish for on-shell external fermions, 
and will be discarded in the analysis that follows.

The (dimension-less) 
self-energy-like contribution
$ \frac{1}{2}\, V_{\alpha\rho}^{{\rm P}\,(1)}(q)$, 
together with another such contribution arising from the 
mirror vertex (not shown),
after trivial manipulations
gives rise to the dimensionful quantity 
\bea
& & \Pi_{\alpha\beta}^{{\rm P}\,(1)}(q) = q^2 
V_{\alpha\rho}^{{\rm P}\,(1)}(q)P_\beta^\rho(q)=
\Pi^{{\rm P}\,(1)}(q)P_{\alpha\beta}(q),\nonumber \\
& &  
\Pi^{{\rm P}\,(1)}(q)=2q^2\int_{L_1}J(q,k).
\label{PP1}
\eea 
$\Pi_{\alpha\beta}^{{\rm P}\,(1)}(q)$
will be added to the
conventional one-loop gluon two-point function $\g_{A_\alpha
A_\beta}^{(1)}(q)$, 
to give rise to the 
the PT one-loop gluon two-point function 
$\widehat\g_{A_\alpha A_\beta}^{(1)}(q)$ (see Fig.\ref{fig2}):
\be
\widehat\g_{A_\alpha A_\beta}^{(1)}(q) = \g_{A_\alpha A_\beta}^{(1)}(q)
+ \Pi_{\alpha\beta}^{{\rm P}\,(1)}(q).
\label{PTprop1}
\ee
Correspondingly, the PT one-loop three-point function 
$\widehat\g^{(1)}_ {A_\alpha\bar\psi\psi}(Q',Q)$ will be defined as
\bea
\widehat\g^{(1)}_{A_\alpha\bar\psi\psi}(Q',Q) & = &
\widehat\Gamma^{A^2\,(1)}_{A_\alpha\bar\psi\psi}(Q',Q)+
\Gamma^{\bar q q\,(1)}_{A_\alpha\bar\psi\psi} \nonumber \\
& = & 
\g^{(1)}_{A_\alpha\bar\psi\psi}(Q',Q)-
\frac12 V_{\alpha\rho}^{{\rm P}\,(1)}(q) \gamma^{\rho}.
\label{1lPTthreepf}
\eea

\begin{figure}[!t]
\bce
\bpi(0,30)(170,-15)

\Photon(55,0)(78.5,0){1.5}{6}
\PhotonArc(90,0)(10,-8,352){1.5}{12}
\Photon(101.5,0)(125,0){1.5}{6}

\Photon(155,0)(178.5,0){1.5}{6}
\DashCArc(190,0)(11.5,0,360){1}
\DashArrowLine(189.5,11.5)(190.5,11.5){1}
\DashArrowLine(190.5,-11.5)(189.5,-11.5){1}
\Photon(201.5,0)(225,0){1.5}{6}

\Photon(255,0)(278.5,0){1.5}{6}
\PhotonArc(290,0)(10,-8,352){1.5}{12}
\Vertex(301.5,0){1.4}

\Text(-15,0)[l]{\footnotesize{$\widehat\g^{(1)}_{A_\alpha A_\beta}(q)$}$\,
=\ \frac12$}
\Text(139,0)[l]{$\scriptstyle{+}$}
\Text(234,0)[l]{$\scriptstyle{+\ 2}$}
\Text(308,0)[l]{$\scriptstyle{P_{\alpha\beta}(q)}$}
\Text(55,-7)[l]{$\scriptstyle{\alpha}$}
\Text(125,-7)[r]{$\scriptstyle{\beta}$}
\Text(155,-7)[l]{$\scriptstyle{\alpha}$}
\Text(225,-7)[r]{$\scriptstyle{\beta}$}
\Text(91,-25)[c]{$\scriptstyle{\rm (a)}$}
\Text(191,-25)[c]{$\scriptstyle{\rm (b)}$}
\Text(291,-25)[c]{$\scriptstyle{\rm (c)}$}

\epi
\ece
\caption{\label{fig2} 
The diagrammatic representation of the PT two-point function
$\widehat\g_{A_\alpha A_\beta}^{(1)}(q)$ 
as the sum of the conventional two-point
function $\g_{A_\alpha A_\beta}^{(1)}(q)$ given by (a) and (b), and the pinch
contributions coming from the vertices (c).}
\end{figure}

We can then compare these results
with the ones we can get from the BQIs of Eqs.\r{bvsqtwopf} and \r{bvsqthreepf}
found in the previous sections.
At one loop these BQIs read
\bea
& & \g^{(1)}_{\widehat{A}_\alpha\widehat{A}_\beta}(q)=\g^{(1)}_{A_\alpha A_\beta}(q)
+2\g^{(1)}_{\Omega_\alpha A^{*}_\rho}(q)\g^{(0)}_{A^\rho A_\beta}(q), 
\nonumber\\
& & \g^{(1)}_{\widehat A_\alpha\bar\psi\psi}(Q',Q)=
\g^{(1)}_{A_\alpha\bar\psi\psi}(Q',Q)+
\g^{(1)}_{\Omega_\alpha A^*_\rho}(q)
\g^{(0)}_{A^\rho\bar\psi\psi}(Q',Q),
\eea
where in the last equation we factor out a $gT^a$ factor.

From the perturbative expansion of Eq.\r{gpert1}, observing that 
$i\g_{\Omega^a_\mu A^{*,b}_\nu}^{(n)}(q)=
\Pi^{(n)}_{\Omega^a_\mu A^{*,b}_\nu}(q)$, one has
\bce
\bpi(0,40)(20,-20)

\Text(-15,-0.5)[r]{$i\g_{\Omega^a_\alpha A^{*,b}_\rho}^{(1)}(q)\,=$}

\Line(-5,0.75)(10,0.75)
\Line(-5,-0.75)(10,-0.75)
\PhotonArc(30,0)(20,0,180){-1.5}{8.5}
\DashCArc(30,0)(20,180,360){1}
\Line(50,0.75)(65,0.75)
\Line(50,-0.75)(65,-0.75)
\DashArrowLine(29.5,-20)(30.5,-20){1}

\Text(-5,7)[l]{$\scriptstyle{\Omega^a_\alpha}$}
\Text(73,7)[r]{$\scriptstyle{A^{*,b}_\rho}$}
\Text(76,-0.5)[l]{$=\,i\delta^{ab}\g_{\Omega_\alpha A^{*}_\rho}^{(1)}(q)$}

\epi
\ece
Therefore, using the Feynman rules of Fig.\ref{fig0}, we find
\bea
\g_{\Omega_\alpha A^*_\rho}^{(1)}(q)&=&
ig_{\alpha\rho}\int_{L_1}J(q,k)\nonumber \\
&=&\frac i2
V^{{\rm P}\,(1)}_{\alpha\rho}(q).
\label{G1vsA1}
\eea
Thus one has the results
\bea
& & 2\g^{(1)}_{\Omega_\alpha A^{*}_\rho}(q)\g^{(0)}_{A^\rho A_\beta}(q)=
\Pi^{{\rm P}\,(1)}_{\alpha\beta}(q), \nonumber\\
& & \g_{\Omega_\alpha A^*_\rho}^{(1)}(q)\g^{(0)}_{A^\rho\bar\psi\psi}(Q',Q)=
-\frac12 V_{\alpha\rho}^{{\rm P}\,(1)}(q)\gamma^\rho,\nonumber
\eea
which will in turn automatically enforce the identifications 
\bea
& & \widehat\g^{(1)}_{A_\alpha A_\beta}(q) \equiv 
\g^{(1)}_{\widehat A_\alpha \widehat A_\beta}(q),\nonumber \\
& & \widehat\g_{A_\alpha\bar\psi\psi}^{(1)}(Q',Q)\equiv
\g_{\widehat A_\alpha\bar\psi\psi}^{(1)}(Q',Q).
\eea

\subsection{Universality (process-independence) of the PT algorithm}

One important question has been whether the construction
of off-shell Green's functions, such as an effective
gluon self-energy, depends on the 
kind of external particles
chosen. This question was settled in 
\cite{Watson:1995tn} by means of detailed 
calculations. In particular it has been shown that 
at one-loop the gluon self-energy constructed by resorting to  
the PT algorithm is universal, in the sense that 
it does not depend on the specific process where it is
embedded. In this subsection we will show how one can arrive at
this result with the aid of the BQIs appearing in Eqs.\r{tgBQI}
and~\r{fgBQI}. 

We will construct  
$\widehat\g_{A_\alpha A_\beta}^{(1)}(q)$ by considering the
process 
$g^{c_1}_{\rho_1} (p_1) 
g^{c_2}_{\rho_2} (p_2)\to g^{c_3}_{\rho_3}(p_3) 
g^{c_4}_{\rho_4}(p_4)$, 
where the $g^{c_i}_{\rho_i}(p_i)$ represent
on-shell gluons, {\it i.e.}, with $p_i^2=0$ and 
$p_i^{\rho_i}\epsilon_{\rho_i}(p_i) = 0$.   

\begin{figure}[!t]
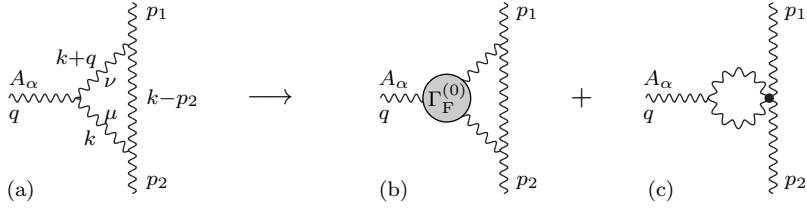

\bce
\bpi(0,50)(170,-20)

\Photon(15,0)(40,0){1.5}{6}
\Photon(60,20)(40,0){-1.5}{6}
\Photon(60,-20)(40,0){1.5}{6}
\Photon(61.5,36)(61.5,-36){1.5}{16.5}

\Photon(155,0)(180,0){1.5}{6}
\Photon(200,20)(180,0){-1.5}{6}
\Photon(200,-20)(180,0){1.5}{6}
\Photon(201.5,36)(201.5,-36){1.5}{16.5}
\GCirc(180,0){9}{0.8}

\Photon(255,0)(278.5,0){1.5}{6}
\PhotonArc(290,0)(10,-8,352){1.5}{12}
\Photon(303,36)(303,-36){-1.5}{16.5}
\Vertex(301.5,0){1.8}

\Text(105,0)[l]{$\longrightarrow$}
\Text(181,0)[c]{$\scriptstyle{\Gamma^{(0)}_{\rm F}}$}
\Text(227.5,0)[l]{$+$}
\Text(15,7)[l]{$\scriptstyle{A_\alpha}$}
\Text(15,-7)[l]{$\scriptstyle{q}$}
\Text(51,6)[l]{$\scriptstyle{\nu}$}
\Text(51,-7)[l]{$\scriptstyle{\mu}$}
\Text(48,15)[r]{$\scriptstyle{k+q}$}
\Text(48,-15)[r]{$\scriptstyle{k}$}
\Text(67,0)[l]{$\scriptstyle{k-p_2}$}
\Text(67,32)[l]{$\scriptstyle{p_1}$}
\Text(67,-32)[l]{$\scriptstyle{p_2}$}
\Text(155,7)[l]{$\scriptstyle{A_\alpha}$}
\Text(155,-7)[l]{$\scriptstyle{q}$}
\Text(207,32)[l]{$\scriptstyle{p_1}$}
\Text(207,-32)[l]{$\scriptstyle{p_2}$}
\Text(255,7)[l]{$\scriptstyle{A_\alpha}$}
\Text(255,-7)[l]{$\scriptstyle{q}$}
\Text(308.5,32)[l]{$\scriptstyle{p_1}$}
\Text(308.5,-32)[l]{$\scriptstyle{p_2}$}
\Text(14,-35)[l]{\scriptsize{(a)}}
\Text(155,-35)[l]{\scriptsize{(b)}}
\Text(256.5,-35)[l]{\scriptsize{(c)}}

\epi
\ece

\caption{\label{un1} The result of carrying out 
the PT decomposition on the three-point function
$\g^{A^2\,(1)}_{A_\alpha A_\beta A_\gamma}(p_1,p_2)$ for the case of 
two external on-shell gluons.
Notice that in graph (c), despite appearances, the vertex 
connecting the loop to the external gluons 
is a tree- and not a four-gluon vertex.}
\end{figure}

The PT algorithm in this case amounts to
carrying out the characteristic three-gluon vertex
decomposition of  Eq.\r{decomp} to the 
graphs contributing to $\g_{A_\alpha^a A_{\rho_1}^{c_1} 
A_{\rho_2}^{c_2}}^{(1)}(q,p_1,p_2)$, which have an external
three-gluon vertex; there are two such graphs, 
out of which only that of Fig.\ref{un1}a gives rise to a propagator-like
contribution. In particular,
the longitudinal momenta $k_{\mu}$ and $(k+q)_{\nu}$
appearing in 
$\Gamma_{\alpha \mu \nu}^{{\rm P}}(q,k,-k-q)$
will be contracted with the corresponding 
three-gluon vertex where one of the two on-shell gluons
is entering ($g^{c_1}_{\rho_1} (p_1)$ and $g^{c_2}_{\rho_2} (p_2)$,
respectively) triggering the tree-level WI of 
Eq.\r{sti3gv0}, which is the exact analogue of 
Eq.\r{PTWI} in the case when the external particles are gluons instead
of quarks. 
It is straightforward to verify 
that again the internal gluon propagator of momentum $k-p_2$
will be canceled by the corresponding piece stemming from the
WI of Eq.\r{sti3gv0}, giving rise to the propagator-like  
diagram of Fig.\ref{un1}c. This piece is simply given by  
(after the  standard insertion of 
$d(q)d^{-1}(q)$ and use of the on-shell conditions)  
\be
\Gamma_{\beta \rho_3 \rho_4}^{(0)}(q,p_3,p_4)\, d(q)  
\bigg[\frac{1}{2}\, \Pi_{\beta\alpha}^{{\rm P}\,(1)}(q)\bigg]\, d(q)
\Gamma_{\alpha \rho_1 \rho_2}^{(0)}(q,p_1,p_2) .
\label{UN1}
\ee

After multiplication by a factor of 2 to take
into account the mirror graphs (not shown)
the above contribution is added to the usual propagator
contributions, also sandwiched between 
$\Gamma_{\beta \rho_3 \rho_4}^{(0)}(q,p_3,p_4)$
and $\Gamma_{\alpha \rho_1 \rho_2}^{(0)}(q,p_1,p_2)$ to give rise
to  the $\widehat\g_{A_\alpha A_\beta}^{(1)}(q)$
of Eq.\r{PTprop1}. A straightforward algebraic manipulation 
of the remaining terms stemming from the WI shows that they
either vanish on-shell, or they combine with the
rest of the diagrams (not shown) to give rise precisely
to the one-loop vertex 
$\g_{\widehat{A}_\alpha^a A_{\rho_1}^{c_1} 
A_{\rho_2}^{c_2}}^{(1)}(q,p_1,p_2)$. 
Of course, in the light of Eq.\r{tgBQI} this is exactly what one 
should obtain, since the subtraction from 
$\g_{A_\alpha^a A_{\rho_1}^{c_1} A_{\rho_2}^{c_2}}^{(1)}(q,p_1,p_2)$ 
of the term given in Eq.\r{UN1}, is nothing but  
$\g_{\widehat{A}_\alpha^a A_{\rho_1}^{c_1} 
A_{\rho_2}^{c_2}}^{(1)}(q,p_1,p_2)$.

We next turn to the slightly more involved case of 
constructing  
$\widehat\g_{A_\alpha A_\beta}^{(1)}(q)$ by embedding it in the
process 
$g^{c_1}_{\rho_1} (p_1) 
g^{c_2}_{\rho_2} (p_2) g^{c_3}_{\rho_3}(p_3)
\to g^{c_4}_{\rho_3}(p_4) 
g^{c_5}_{\rho_5}(p_5) g^{c_6}_{\rho_6}(p_6)$, 
where, as before, the $g^{c_i}_{\rho_i}(p_i)$ represent
on-shell gluons, with $p_i^2=0$ and 
$p_i^{\rho_i}\epsilon_{\rho_i}(p_i) = 0$.   

\begin{figure}[!t]
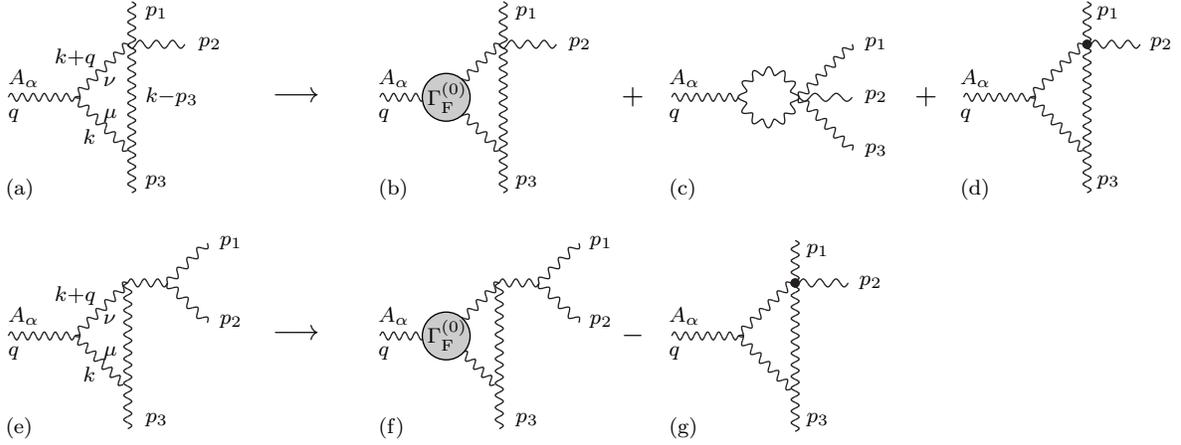

\bce
\bpi(0,170)(230,-115)

\Photon(15,0)(40,0){1.5}{6}
\Photon(60,20)(40,0){-1.5}{6}
\Photon(60,-20)(40,0){1.5}{6}
\Photon(61.5,36)(61.5,-36){1.5}{16.5}
\Photon(60,20)(81.5,20){-1.5}{3.5}

\Photon(155,0)(180,0){1.5}{6}
\Photon(200,20)(180,0){-1.5}{6}
\Photon(200,-20)(180,0){1.5}{6}
\Photon(201.5,36)(201.5,-36){1.5}{16.5}
\Photon(200,20)(221.5,20){-1.5}{3.5}
\GCirc(180,0){9}{0.8}

\Photon(265,0)(290,0){1.5}{6}
\PhotonArc(301.5,0)(10,-8,352){1.5}{12}
\Photon(313,0)(333,20){1.5}{6}
\Photon(313,0)(333,-20){-1.5}{6}
\Photon(313,0)(333,0){-1.5}{2.5}

\Photon(375,0)(400,0){1.5}{6}
\Photon(420,20)(400,0){-1.5}{6}
\Photon(420,-20)(400,0){1.5}{6}
\Photon(421.5,36)(421.5,-36){1.5}{16.5}
\Photon(420,20)(441.5,20){-1.5}{3.5}
\Vertex(421.5,20){1.8}

\Text(115,0)[l]{$\longrightarrow$}
\Text(181,0)[c]{$\scriptstyle{\Gamma^{(0)}_{\rm F}}$}
\Text(247,0)[l]{$+$}
\Text(15,7)[l]{$\scriptstyle{A_\alpha}$}
\Text(15,-7)[l]{$\scriptstyle{q}$}
\Text(51,6)[l]{$\scriptstyle{\nu}$}
\Text(51,-7)[l]{$\scriptstyle{\mu}$}
\Text(48,15)[r]{$\scriptstyle{k+q}$}
\Text(48,-15)[r]{$\scriptstyle{k}$}
\Text(67,0)[l]{$\scriptstyle{k-p_3}$}
\Text(67,32)[l]{$\scriptstyle{p_1}$}
\Text(87,20)[l]{$\scriptstyle{p_2}$}
\Text(67,-32)[l]{$\scriptstyle{p_3}$}

\Text(155,7)[l]{$\scriptstyle{A_\alpha}$}
\Text(155,-7)[l]{$\scriptstyle{q}$}
\Text(207,32)[l]{$\scriptstyle{p_1}$}
\Text(227,20)[l]{$\scriptstyle{p_2}$}
\Text(207,-32)[l]{$\scriptstyle{p_3}$}

\Text(265,7)[l]{$\scriptstyle{A_\alpha}$}
\Text(265,-7)[l]{$\scriptstyle{q}$}
\Text(339,20)[l]{$\scriptstyle{p_1}$}
\Text(339,0)[l]{$\scriptstyle{p_2}$}
\Text(339,-20)[l]{$\scriptstyle{p_3}$}

\Text(358,0)[l]{$+\ $}

\Text(375,7)[l]{$\scriptstyle{A_\alpha}$}
\Text(375,-7)[l]{$\scriptstyle{q}$}
\Text(427,32)[l]{$\scriptstyle{p_1}$}
\Text(447,20)[l]{$\scriptstyle{p_2}$}
\Text(427,-32)[l]{$\scriptstyle{p_3}$}

\Text(14,-35)[l]{\scriptsize{(a)}}
\Text(14,-125)[l]{\scriptsize{(e)}}
\Text(155,-35)[l]{\scriptsize{(b)}}
\Text(155,-125)[l]{\scriptsize{(f)}}
\Text(265,-35)[l]{\scriptsize{(c)}}
\Text(265,-125)[l]{\scriptsize{(g)}}
\Text(375,-35)[l]{\scriptsize{(d)}}

\Photon(15,-90)(40,-90){1.5}{6}
\Photon(60,-70)(40,-90){-1.5}{6}
\Photon(58.3,-108.3)(40,-90){1.5}{6}
\Photon(60,-70)(60,-125){-1.5}{12}
\Photon(60,-70)(75,-70){1.5}{3}
\Photon(75,-70)(90,-55){1.5}{4}
\Photon(75,-70)(90,-85){-1.5}{4}

\Photon(155,-90)(180,-90){1.5}{6}
\Photon(200,-70)(180,-90){-1.5}{6}
\Photon(198.3,-108.3)(180,-90){1.5}{6}
\Photon(200,-70)(200,-125){-1.5}{12}
\Photon(200,-70)(215,-70){1.5}{3}
\Photon(215,-70)(230,-55){1.5}{4}
\Photon(215,-70)(230,-85){-1.5}{4}
\GCirc(180,-90){9}{0.8}

\Photon(265,-90)(290,-90){1.5}{6}
\Photon(310,-70)(290,-90){-1.5}{6}
\Photon(310,-110)(290,-90){1.5}{6}
\Photon(311.5,-54)(311.5,-126){1.5}{16.5}
\Photon(310,-70)(331.5,-70){-1.5}{3.5}
\Vertex(311.5,-70){1.8}

\Text(265,-83)[l]{$\scriptstyle{A_\alpha}$}
\Text(265,-97)[l]{$\scriptstyle{q}$}
\Text(317,-58)[l]{$\scriptstyle{p_1}$}
\Text(337,-70)[l]{$\scriptstyle{p_2}$}
\Text(317,-122)[l]{$\scriptstyle{p_3}$}

\Text(115,-90)[l]{$\longrightarrow$}
\Text(181,-90)[c]{$\scriptstyle{\Gamma^{(0)}_{\rm F}}$}
\Text(247,-90)[l]{$-$}
\Text(15,-83)[l]{$\scriptstyle{A_\alpha}$}
\Text(15,-97)[l]{$\scriptstyle{q}$}
\Text(51,-84)[l]{$\scriptstyle{\nu}$}
\Text(51,-97)[l]{$\scriptstyle{\mu}$}
\Text(48,-75)[r]{$\scriptstyle{k+q}$}
\Text(48,-105)[r]{$\scriptstyle{k}$}
\Text(95,-55)[l]{$\scriptstyle{p_1}$}
\Text(95,-85)[l]{$\scriptstyle{p_2}$}
\Text(67,-122)[l]{$\scriptstyle{p_3}$}

\Text(155,-83)[l]{$\scriptstyle{A_\alpha}$}
\Text(155,-97)[l]{$\scriptstyle{q}$}
\Text(235,-55)[l]{$\scriptstyle{p_1}$}
\Text(235,-85)[l]{$\scriptstyle{p_2}$}
\Text(207,-122)[l]{$\scriptstyle{p_3}$}

\epi
\ece
\caption{\label{un2} The result of carrying out 
the PT decomposition on the 1PI four-point function 
$\g^{A^2\,(1)}_{A_\alpha A_\beta A_\gamma A_\delta}(p_1,p_2,p_3)$ 
(a) and the 1PR 
four-point function (b) for the case of three external 
on-shell gluons (permutations are not shown). 
Again, despite appearances, the vertex 
connecting the loop to the external gluons in diagrams (d) and (g), 
is a tree-gluon and not a four-gluon vertex.}
\end{figure}

As before, one should carry out the characteristic three-gluon vertex
decomposition of  Eq.\r{decomp} to the 
graphs which have an external
three-gluon vertex; there are various such graphs,
but the essence of the relevant rearrangements 
can be captured by looking at the graphs shown in Fig.\ref{un2}.
The graph of Fig.\ref{un2}a contributes to the
1PI one-loop four-gluon vertex $\g_{A_\alpha^a A_{\rho_1}^{c_1} 
A_{\rho_2}^{c_2} A_{\rho_3}^{c_3}}^{(1)}(q,p_1,p_2,p_3)$,
whereas the graph of Fig.\ref{un2}e is 1PR and contributes to the
one-loop three-gluon vertex nested 
inside the process we consider. Notice in particular that
unlike the one-loop three-gluon vertex considered 
in the previous process (Fig.\ref{un1}a),
the one appearing in  Fig.\ref{un2}e has not one but two off-shell legs
(those carrying momenta $q$ and $p_1+p_2$). 

The action of the longitudinal (pinching) momenta stemming from
the PT decomposition of the external tree-level three-gluon vertex
appearing in the graph of Fig.\ref{un2}a gives rise to the 
propagator-like contribution of Fig.\ref{un2}c, given by
the same expression as in Eq.\r{UN1}, with the only difference
that the contribution
$d(q) \bigg[\frac{1}{2}\, \Pi_{\beta\alpha}^{{\rm P}\,(1)}(q)\bigg] d(q)$
is sandwiched between two tree-level four-gluon vertices instead
of two  tree-level three-gluon vertices. In addition  
to this propagator-like contribution the pinching momenta
give also rise to
contributions of the type shown in Fig.\ref{un2}d, 
by virtue of the elementary WI
\bea
q_1^{\mu} \g_{A_\mu^a A_\nu^b A_\alpha^c A_\beta^d}^{(0)}(q_1,q_2,q_3,q_4) 
&=&
-f^{abe}\, \g_{A_\alpha^c A_\beta^d A_\nu^e}^{(0)}(q_3,q_4,q_1+q_2)
-f^{ace}\, \g_{A_\beta^d A_\nu^b A_\alpha^e}^{(0)}(q_4,q_2,q_1+q_3)
\nonumber\\
&-& f^{ade} \, \g_{A_\nu^b A_\alpha^c
A_\beta^e}^{(0)}(q_2,q_3,q_1+q_4)
\eea

When the PT decomposition is implemented in the graph of Fig.\ref{un2}e,
it gives rise to various contributions, the most characteristic of 
which are depicted in Fig.\ref{un2}. Most notably, the parts of the WI
which in the three-gluon vertex of the
previous process that we considered were vanishing on-shell, 
because they were proportional to $p_i^2$, now they simply cancel
the off-shell propagator $d(p_1+p_2)$, thus giving rise to the
effectively 1PI graph shown in Fig.\ref{un2}g. This latter 
contribution will cancel exactly against the one shown in Fig.\ref{un2}d. 
It is important to notice at this point  
how all the above cancellations are encoded in the identities of  
Eqs.\r{tgBQIOS} and~\r{fgBQI}. In particular, the three last terms
appearing on the RHS of Eq.\r{fgBQI} are nothing but 
the terms collectively depicted in Fig.\ref{un2}d, together 
with all the relevant permutations (not shown).  
Similarly, the two last terms on the RHS of Eq.\r{tgBQIOS}
are precisely the terms shown in Fig.\ref{un2}d, which 
would have vanished if the external legs had been on-shell
[as happens in Eq.\r{tgBQI}].   
Notice that 
all these terms are proportional to the same basic quantity, 
namely the three-point function $\g_{\Omega A^{*} A}^{(1)}(q_1,q_2)$. 

\section{\label{sec:four} Two-loop case revisited}

\begin{figure}[!t]
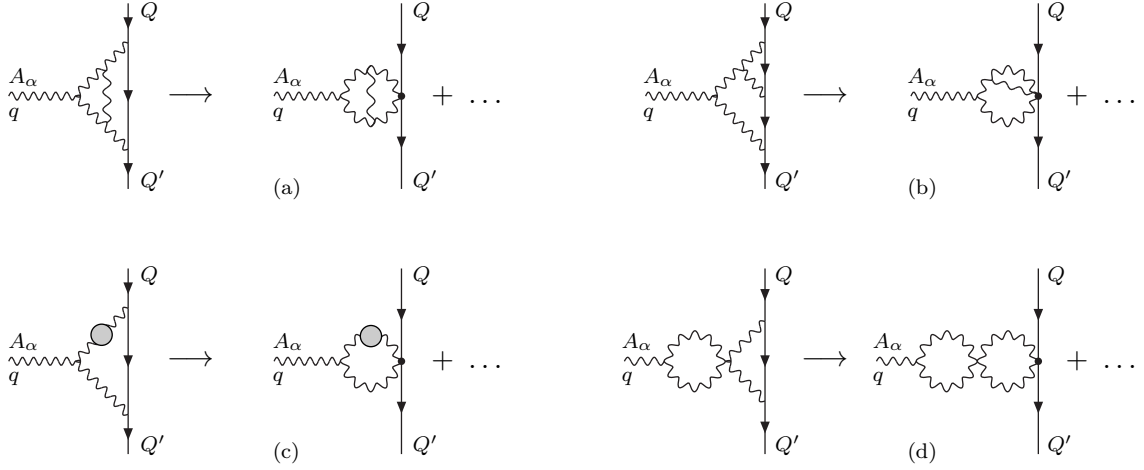

\bce
\bpi(0,150)(225,-120)

\Photon(15,0)(40,0){1.5}{6}
\Photon(60,20)(40,0){-1.5}{6}
\Photon(60,-20)(40,0){1.5}{6}
\Photon(52,9.8)(52,-10){-1.5}{3}
\ArrowLine(60,35)(60,20)
\ArrowLine(60,20)(60,-20)
\ArrowLine(60,-20)(60,-35)

\Photon(115,0)(140,0){1.5}{6}
\PhotonArc(151.5,0)(10,-8,352){1.5}{12}
\Photon(151.5,-11.5)(151.5,11.5){-1.5}{3}
\ArrowLine(163,35)(163,0)
\ArrowLine(163,0)(163,-35)
\Vertex(163,0){1.4}

\Photon(255,0)(280,0){1.5}{6}
\Photon(300,20)(280,0){-1.5}{6}
\Photon(300,-20)(280,0){1.5}{6}
\Photon(300,0)(291.5,9.5){1.5}{3}
\ArrowLine(300,35)(300,20)
\ArrowLine(300,-20)(300,-35)
\ArrowLine(300,20)(300,0)
\ArrowLine(300,0)(300,-20)

\Photon(355,0)(380,0){1.5}{6}
\PhotonArc(391.5,0)(10,-8,352){1.5}{12}
\Photon(403,0)(385.7,6){-1.0}{3}
\ArrowLine(403,35)(403,0)
\ArrowLine(403,0)(403,-35)
\Vertex(403,0){1.4}

\Text(75,0)[l]{$\longrightarrow$}
\Text(175,0)[l]{$+\ \dots$}
\Text(315,0)[l]{$\longrightarrow$}
\Text(415,0)[l]{$+\ \dots$}
\Text(15,7)[l]{$\scriptstyle{A_\alpha}$}
\Text(15,-7)[l]{$\scriptstyle{q}$}
\Text(65,32)[l]{$\scriptstyle{Q}$}
\Text(65,-32)[l]{$\scriptstyle{Q'}$}
\Text(115,7)[l]{$\scriptstyle{A_\alpha}$}
\Text(115,-7)[l]{$\scriptstyle{q}$}
\Text(168,32)[l]{$\scriptstyle{Q}$}
\Text(168,-32)[l]{$\scriptstyle{Q'}$}
\Text(255,7)[l]{$\scriptstyle{A_\alpha}$}
\Text(255,-7)[l]{$\scriptstyle{q}$}
\Text(305,32)[l]{$\scriptstyle{Q}$}
\Text(305,-32)[l]{$\scriptstyle{Q'}$}
\Text(355,7)[l]{$\scriptstyle{A_\alpha}$}
\Text(355,-7)[l]{$\scriptstyle{q}$}
\Text(408,32)[l]{$\scriptstyle{Q}$}
\Text(408,-32)[l]{$\scriptstyle{Q'}$}
\Text(115,-35)[l]{\scriptsize{(a)}}
\Text(355,-35)[l]{\scriptsize{(b)}}

\Photon(15,-100)(40,-100){1.5}{6}
\Photon(60,-80)(40,-100){-1.5}{6}
\Photon(60,-120)(40,-100){1.5}{6}
\ArrowLine(60,-65)(60,-80)
\ArrowLine(60,-80)(60,-120)
\ArrowLine(60,-120)(60,-135)
\GCirc(50,-90){4}{0.8}

\Photon(115,-100)(140,-100){1.5}{6}
\PhotonArc(151.5,-100)(10,-8,352){1.5}{12}
\ArrowLine(163,-65)(163,-100)
\ArrowLine(163,-100)(163,-135)
\GCirc(151.5,-90.5){4}{0.8}
\Vertex(163,-100){1.4}

\Photon(262,-100)(247,-100){1.5}{3}
\Photon(285,-100)(300,-85){-1.5}{4}
\Photon(285,-100)(300,-115){1.5}{4}
\PhotonArc(273.5,-100)(10,-8,352){1.5}{12}
\ArrowLine(300,-65)(300,-85)
\ArrowLine(300,-85)(300,-115)
\ArrowLine(300,-115)(300,-135)

\Photon(357,-100)(342,-100){1.5}{3}
\PhotonArc(368.5,-100)(10,-8,352){1.5}{12}
\PhotonArc(391.5,-100)(10,-8,352){1.5}{12}
\ArrowLine(403,-65)(403,-100)
\ArrowLine(403,-100)(403,-135)
\Vertex(403,-100){1.4}

\Text(75,-100)[l]{$\longrightarrow$}
\Text(175,-100)[l]{$+\ \dots$}
\Text(315,-100)[l]{$\longrightarrow$}
\Text(415,-100)[l]{$+\ \dots$}
\Text(15,-93)[l]{$\scriptstyle{A_\alpha}$}
\Text(15,-107)[l]{$\scriptstyle{q}$}
\Text(65,-68)[l]{$\scriptstyle{Q}$}
\Text(65,-132)[l]{$\scriptstyle{Q'}$}
\Text(115,-93)[l]{$\scriptstyle{A_\alpha}$}
\Text(115,-107)[l]{$\scriptstyle{q}$}
\Text(168,-68)[l]{$\scriptstyle{Q}$}
\Text(168,-132)[l]{$\scriptstyle{Q'}$}
\Text(247,-93)[l]{$\scriptstyle{A_\alpha}$}
\Text(247,-107)[l]{$\scriptstyle{q}$}
\Text(305,-68)[l]{$\scriptstyle{Q}$}
\Text(305,-132)[l]{$\scriptstyle{Q'}$}
\Text(342,-93)[l]{$\scriptstyle{A_\alpha}$}
\Text(342,-107)[l]{$\scriptstyle{q}$}
\Text(408,-68)[l]{$\scriptstyle{Q}$}
\Text(408,-132)[l]{$\scriptstyle{Q'}$}
\Text(115,-135)[l]{\scriptsize{(c)}}
\Text(355,-135)[l]{\scriptsize{(d)}}

\epi
\ece

\caption{\label{fig2.1} Enforcing the PT decomposition on the 
three-point function $\g^{A^2\,(2)}_{A\bar\psi\psi}(Q',Q)$, gives rise to the 
topologies $I_1$ (a), $I_3$ (b), $I_4$ (c) and $I_2$ (d) of Eq.\r{2lptcont}.
The ellipses represents terms that whether they cancel or they modify 
the ghost structure.}
\end{figure}

At the two-loop level \cite{Papavassiliou:1999az,Papavassiliou:1999bb}
we start again 
by carrying out the decomposition Eq.\r{deco1} of the two-loop
three-point function $\g^{(2)}_{A_\alpha\bar\psi\psi}(Q',Q)$ ;  
now all four categories of diagrams appearing on the RHS of 
Eq.(\ref{deco1})
are non-vanishing. Next (see Fig.\ref{fig2.1}), we implement the vertex
decomposition Eq.\r{decomp} inside the
$\Gamma_{A_\alpha\bar\psi\psi}^{A^2\,(2)}(Q',Q)$ part, which will again trigger
elementary WIs, leading us to the result
\be
\Gamma_{A_\alpha\bar\psi\psi}^{A^2\,(2)}(Q',Q)=
\widehat{\Gamma}_{A_\alpha\bar\psi\psi}^{A^2\,(2)}(Q',Q)
+\frac{1}{2}V_{\alpha\rho}^{{\rm P}\,(2)}(q) \gamma^{\rho}
+\frac{1}{2} F^{{\rm P}\,(2)}_{\alpha}(Q,Q'),
\ee
with 
\bea
V_{\alpha\rho}^{{\rm P}\,(2)}(q) & = & I_4 L_{\alpha\rho}(\ell,k)
+ (2 I_{2} + I_{3}) g_{\alpha\rho} \nonumber \\
& - & I_1 \left[k_{\rho}g_{\alpha\sigma}+  
\Gamma_{\sigma\rho\alpha}^{(0)}(-k,-\ell,k+\ell)\right](\ell-q)^{\sigma},
\nonumber \\
F^{{\rm P}\,(2)}_{\alpha}(Q',Q) &=&
d(q)\Pi_{\alpha}^{{\rm P}\,(1)\beta}(q)
\widehat\g_{A_\beta\bar\psi\psi}^{(1)}(Q',Q) + 
Y_{\alpha}^{{\rm P}\,(2)}(Q',Q),\nonumber \\
Y_{{\rm P}\,\alpha}^{(2)}(Q',Q) & = & 
X_{1\,\alpha}^{(1)}(Q',Q)\Sigma^{(1)}(Q')+
\Sigma^{(1)}(Q)X_{2\,\alpha}^{(1)}(Q',Q).
\label{2lptcont}
\eea
The integrals $I_i$ appearing in Eq.\r{2lptcont} are defined as
\bea
iI_{1} &=& g^4 C_A^2 \int_{L_2}
[\ell^2 (\ell-q)^2 k^2 (k+\ell)^2 (k+\ell-q)^2]^{-1},\nonumber\\
iI_{2}  &=& g^4 C_A^2 \int_{L_2}
[\ell^2 (\ell-q)^2 k^2 (k+q)^2]^{-1},\nonumber\\
iI_{3} &=& g^4 C_A^2 \int_{L_2}
[\ell^2 (\ell-q)^2 k^2 (k+\ell)^2]^{-1},\nonumber\\
iI_{4} &=& g^4 C_A^2 \int_{L_2}
[\ell^2 \ell^2 (\ell-q)^2 k^2 (k+\ell)^2]^{-1},
\label{2LInt}
\eea
where we have defined the (two-loop) integral measure
\be
\int_{L_2}\equiv\left(\mu^{2\varepsilon}\right)^2
\int\!\frac{d^dk}{(2\pi)^d}\int\!\frac{d^d\ell}{(2\pi)^d}.
\ee

As before the term $Y_\alpha^{{\rm P}\,(2)}(Q',Q)$ will vanish for on-shell
external fermions so that it will be omitted all together.
The term $\frac12V^{{\rm P}\,(2)}_{\alpha\rho}(q)\gamma^\rho$ represents the
total propagator-like term originating from the two-loop three-point function
$\g_{A_\alpha\bar\psi\psi}^{(2)}(Q',Q)$: together with the equal contribution
coming from the mirror set of two-loop vertex diagrams, will give rise to the
self-energy term
\be
\Pi_{\alpha\beta}^{{\rm P}\,(2)}(q) = q^2V^{{\rm P}\,(2)}_{\alpha\rho}(q)
P^\rho_\beta(q),
\ee
which will be part of the two-loop PT gluon two-point function.

However beyond one-loop, 
this is not the end of the story, since one has to take
into account the conversion of 1PR strings of conventional two-point functions
$\g_{AA}(q)$, into strings containing PT 
two-point functions $\widehat\g_{AA}(q)$
\cite{Papavassiliou:1995fq,Papavassiliou:1996gs}. 
Actually the term $\frac12F^{{\rm P}\,(2)}_{\alpha}(Q',Q)$ appearing in
Eq.\r{2lptcont} is half of the vertex-like necessary to cancel the corresponding
term appearing during the aforementioned conversion (the other half will come
from the mirror set of diagrams).

The PT two-loop two-point function is then given by 
\be
{\widehat\g}^{(2)}_{A_\alpha A_\beta}(q) =
\g^{(2)}_{A_\alpha A_\beta}(q) + \Pi_{\alpha\beta}^{{\rm P}\,(2)}(q) -
R^{{\rm P}\,(2)}_{\alpha\beta}(q).
\label{2lPTtwopf}
\ee
where $R^{{\rm P}\,(2)}_{\alpha\beta}(q)$, which also stems from the conversion
of the conventional 1PR string into a 1PR PT one, is given by
\be
iR^{{\rm P}\,(2)}_{\alpha\beta}(q)=
\g^{(1)}_{A_\alpha A_\rho}(q) V_{\beta}^{{\rm P}\,(1)\,\rho}(q) + 
\frac{3}{4} \,V_{\alpha\rho}^{{\rm P}\,(1)}(q)V_{
\beta}^{{\rm P}\,(1)\,\rho}(q).
\ee
Correspondingly, the two-loop PT three-point function
$\widehat\g^{(2)}_{A_\alpha\bar\psi\psi}(Q',Q)$ will be defined as
\bea
\widehat\g^{(2)}_{A_\alpha\bar\psi\psi}(Q',Q) & = &
\widehat\Gamma^{A^2\,(2)}_{A_\alpha\bar\psi\psi}(Q',Q)+
\Gamma^{\bar c c\,(2)}_{A_\alpha\bar\psi\psi}(Q',Q)+
\Gamma^{A^3\,(2)}_{A_\alpha\bar\psi\psi}(Q',Q)+
\Gamma^{\bar q q\,(2)}_{A_\alpha\bar\psi\psi}(Q',Q) \nonumber \\
& = & \g^{(2)}_{A_\alpha\bar\psi\psi}(Q',Q)+\frac12 V^{{\rm
P}(2)}_{\alpha\rho}(q)\gamma^\rho+\frac 12d(q)
\Pi^{{\rm P}\,(1)\,\rho}_{\alpha}(q)
\widehat\g_{A_\rho\bar\psi\psi}^{(1)}(Q',Q).
\label{2lPTthreepf}
\eea

We can now compare these results with the one coming from the BQIs, 
so that we will be able to verify directly that
\bea
& & \widehat\g^{(2)}_{A_\alpha A_\beta}(q)\equiv
\g^{(2)}_{\widehat A_\alpha \widehat A_\beta}(q), \label{PTvsBFM2ltwopf} \\
& & \widehat\g^{(2)}_{A_\alpha\bar\psi\psi}(Q',Q)\equiv
\g^{(2)}_{\widehat A_\alpha\bar\psi\psi}(Q',Q).
\label{PTvsBFM2lthreepf}
\eea
 
At the two-loop level the BQIs of Eqs.\r{bvsqtwopf} and \r{bvsqthreepf} read
\bea
\g^{(2)}_{\widehat{A}_\alpha\widehat{A}_\beta}(q) & = & \g^{(2)}_{A_\alpha
A_\beta}(q)
+2\g^{(2)}_{\Omega_\beta A^{*}_\rho}(q)\g^{(0)}_{A^\rho A_\alpha}(q)
+2\g^{(1)}_{\Omega_\beta A^{*}_\rho}(q)\g^{(1)}_{A^\rho A_\alpha}(q)
\nonumber \\
& + & \g^{(1)}_{\Omega_\alpha A^*_\rho}(q)\g^{(0)}_{A^\rho A^\sigma}(q)
\g^{(1)}_{\Omega_\beta A^*_\sigma}(q),
\label{bvsqtwopf2l}\\
\g^{(2)}_{\widehat A_\alpha\bar\psi\psi}(Q',Q) & = &
\g^{(2)}_{A_\alpha\bar\psi\psi}(Q',Q)+
\g^{(2)}_{\Omega_\alpha A^{*}_\rho}(q)
\g^{(0)}_{A^\rho\bar\psi\psi}(Q',Q) \nonumber \\
& + &
\g^{(1)}_{\Omega_\alpha A^{*}_\rho}(q)
\g^{(1)}_{A^\rho\bar\psi\psi}(Q',Q).
\label{bvsqthreepf2l}
\eea
Consider then Eq.\r{bvsqtwopf2l}. We will now prove that
\be
\Pi^{{\rm P}\,(2)}_{\alpha\beta}(q)-R^{{\rm P}\,(2)}_{\alpha\beta}(q)=
2\g^{(2)}_{\Omega_\beta A^{*}_\rho}(q)\g^{(0)}_{A^\rho A_\alpha}(q)
+2\g^{(1)}_{\Omega_\beta A^{*}_\rho}(q)\g^{(1)}_{A^\rho A_\alpha}(q)
+\left[\g^{(1)}_{\Omega A^*}(q)\right]^2\g^{(0)}_{A_\alpha A_\beta}(q).
\ee
To this end we notice that the left-hand side above can be written as
\bea
\Pi^{{\rm P}\,(2)}_{\alpha\beta}(q)-
R^{{\rm P}\,(2)}_{\alpha\beta}(q) & = & q^2P_\beta^\rho(q)
\bigg\{I_4 L_{\alpha\rho}(\ell,k)
+ I_{3}g_{\alpha\rho}  \nonumber \\
& - & 
I_1 \left[k_{\rho}g_{\alpha\sigma}+  
\Gamma_{\sigma\rho\alpha}^{(0)}(-k,-\ell,k+\ell)\right](\ell-q)^{\sigma}\bigg\}
\nonumber \\
&+& iV^{{\rm
P}\,(1)\,\rho}_\beta(q)\g^{(1)}_{A_\rho A_\alpha}(q)-q^2I_2P_{\alpha\beta}(q).
\label{ta}
\eea
Then Eq.\r{G1vsA1} implies
\bea
& & 2\g^{(1)}_{\Omega_\beta A^{*}_\rho}(q)\g^{(1)}_{A^\rho A_\alpha}(q)=
iV^{{\rm
P}\,(1)\,\rho}_\beta(q)\g^{(1)}_{A_\rho A_\alpha}(q),\nonumber \\
& & \g^{(1)}_{\Omega_\alpha A^*_\rho}(q)\g^{(0)}_{A^\rho A^\sigma}(q)
\g^{(1)}_{\Omega_\beta A^*_\sigma}(q)=
-q^2I_2P_{\alpha\beta}(q).
\eea
Finally, from the perturbative expansion \r{gpert1} one has that
\bce
\bpi(0,65)(140,-35)

\Text(-17,-0.5)[r]{$i\g^{(2)}_{\Omega_\alpha A^*_\rho}(q)\,=$}

\Line(-5,0.75)(10,0.75)
\Line(-5,-0.75)(10,-0.75)
\PhotonArc(30,0)(20,0,180){-1.5}{8.5}
\DashCArc(30,0)(20,180,360){1}
\Line(50,0.75)(65,0.75)
\Line(50,-0.75)(65,-0.75)
\Photon(30,-20)(30,18.5){1.5}{6}
\DashArrowLine(13,-10)(13.6,-11){1}
\DashArrowLine(46,-12)(46.7,-11){1}

\Line(90,0.75)(105,0.75)
\Line(90,-0.75)(105,-0.75)
\PhotonArc(125,0)(20,90,180){-1.5}{4}
\PhotonArc(125,0)(20,270,360){-1.5}{4}
\DashCArc(125,0)(20,0,90){1}
\DashCArc(125,0)(20,180,270){1}
\DashArrowLine(125,-20)(125,20){1}
\Line(145,0.75)(160,0.75)
\Line(145,-0.75)(160,-0.75)
\DashArrowLine(108,-10)(108.6,-11){1}
\DashArrowLine(141,12)(141.7,11){1}

\Line(185,0.75)(200,0.75)
\Line(185,-0.75)(200,-0.75)
\PhotonArc(220,0)(20,0,180){-1.5}{8.5}
\DashCArc(220,0)(20,180,360){1}
\Line(240,0.75)(255,0.75)
\Line(240,-0.75)(255,-0.75)
\GCirc(220,18.5){6}{0.8}
\DashArrowLine(219.5,-20)(220.5,-20){1}

\Line(280,0.75)(295,0.75)
\Line(280,-0.75)(295,-0.75)
\PhotonArc(315,0)(20,0,180){-1.5}{8.5}
\DashCArc(315,0)(20,180,360){1}
\Line(335,0.75)(350,0.75)
\Line(335,-0.75)(350,-0.75)
\GCirc(315,-20){6}{0.8}
\DashArrowLine(298,-10)(298.6,-11){1}
\DashArrowLine(331,-12)(331.7,-11){1}

\Text(77.5,0)[c]{$\scriptstyle{+}$}
\Text(173.5,0)[c]{$\scriptstyle{+}$}
\Text(268.5,0)[c]{$\scriptstyle{+}$}

\Text(-5,-25)[l]{\scriptsize{(a)}}
\Text(90,-25)[l]{\scriptsize{(b)}}
\Text(185,-25)[l]{\scriptsize{(c)}}
\Text(280,-25)[l]{\scriptsize{(d)}}

\Text(-5,7)[l]{$\scriptstyle{\Omega_\alpha}$}
\Text(90,7)[l]{$\scriptstyle{\Omega_\alpha}$}
\Text(185,7)[l]{$\scriptstyle{\Omega_\alpha}$}
\Text(280,7)[l]{$\scriptstyle{\Omega_\alpha}$}

\Text(67,7)[r]{$\scriptstyle{A^*_\rho}$}
\Text(162,7)[r]{$\scriptstyle{A^*_\rho}$}
\Text(257,7)[r]{$\scriptstyle{A^*_\rho}$}
\Text(352,7)[r]{$\scriptstyle{A^*_\rho}$}

\epi
\ece
where the blobs represent one-loop correction to the corresponding propagator.
Using the Feynman rules of Fig.\ref{fig0}, it is then straightforward to establish 
the following identities
\bea
& & {\rm (a)}=\frac12I_1\Gamma^{(0)}_{\sigma\rho\alpha}(-k,-\ell,k+\ell)
(\ell-q)^\sigma, \qquad
{\rm (b)}=\frac12I_1(\ell-q)_\alpha k_\rho, \nonumber\\
& & {\rm (c)}=-\frac12I_4L_{\alpha\rho}(\ell,k), \hspace{3.85cm}
{\rm (d)}=-\frac12I_3g_{\alpha\rho},
\label{li}
\eea
so that
\bea
2\g^{(2)}_{\Omega_\beta A^{*}_\rho}(q)\g^{(0)}_{A^\rho A_\alpha}(q) & = &
q^2P_\beta^\rho(q)
\bigg\{I_4 L_{\alpha\rho}(\ell,k)
+ I_{3}g_{\alpha\rho} \nonumber \\
& - & I_1 \left[k_{\rho}g_{\alpha\sigma}+  
\Gamma_{\sigma\rho\alpha}^{(0)}(-k,-\ell,k+\ell)\right](\ell-q)^{\sigma}
\bigg\}.
\eea
Thus Eq.\r{PTvsBFM2ltwopf} is proved.

Finally consider the two-loop PT three-point function Eq.\r{2lPTthreepf}, which
using  Eq.\r{1lPTthreepf}, reads
\bea
\widehat\g^{(2)}_{A_\alpha\bar\psi\psi}(Q',Q)  & = &
\g^{(2)}_{A_\alpha\bar\psi\psi}(Q',Q)+
\frac i2V_\alpha^{{\rm P}\,(1)\,\rho}(q)\g_{A_\rho\bar\psi\psi}^{(1)}
-\frac12\bigg\{I_4 L_{\alpha\rho}(\ell,k)
+ I_{3}g_{\alpha\rho} \nonumber \\
& - & I_1 \left[k_{\rho}g_{\alpha\sigma}+  
\Gamma_{\sigma\rho\alpha}^{(0)}(-k,-\ell,k+\ell)\right](\ell-q)^{\sigma}\bigg\}
\gamma^\rho.
\eea
Then using Eqs.\r{G1vsA1} and \r{li} we get
\bea
\g^{(1)}_{\Omega_\alpha A^{*}_\rho}(q)
\g^{(1)}_{A^\rho\bar\psi\psi}(Q',Q) & = &
\frac i2V_\alpha^{{\rm P}\,(1)\,\rho}(q)\g_{A_\rho\bar\psi\psi}^{(1)},
\nonumber \\
\g^{(2)}_{\Omega_\alpha A^{*}_\rho}(q)
\g^{(0)}_{A^\rho\bar\psi\psi}(Q',Q) & = &
\frac12\bigg\{I_4 L_{\alpha\rho}(\ell,k)
+ I_{3}g_{\alpha\rho} \nonumber \\
& - & I_1 \left[k_{\rho}g_{\alpha\sigma}+  
\Gamma_{\sigma\rho\alpha}^{(0)}(-k,-\ell,k+\ell)\right](\ell-q)^{\sigma}\bigg\}
\gamma^\rho,
\eea
so that Eq.\r{PTvsBFM2lthreepf} is also proved.

\section{\label{secIP} A New Result: The two-loop intrinsic Pinch technique} 

In the intrinsic PT construction one avoids the embedding of the PT objects
into $S$-matrix elements; of course, all  results of the intrinsic PT are
identical to those obtained in the $S$-matrix PT context.  The basic idea, is
that the pinch graphs, which are essential in  canceling the gauge dependences
of ordinary diagrams, are always missing one or more propagators corresponding
to the external legs of the improper Green's function in question. It then
follows that the gauge-dependent parts of such ordinary diagrams must also be
missing one or more external propagators. Thus the intrinsic PT construction
goal is to isolate systematically the parts of 1PI diagrams that are 
proportional to the inverse propagators of the external legs and simply discard
them.  The important point is that these inverse propagators  arise from the
STI satisfied by the three-gluon vertex appearing inside appropriate sets of
diagrams, when it is contracted by longitudinal momenta. The STI triggered is
nothing but  Eq.(\ref{sti3gv}), {\it i.e.},
\bea
p_1^\mu\g_{A_\alpha A_\mu A_\nu}(q,p_1,p_2) & = & 
\left[i\Delta^{(-1)\,\rho}_{\nu}(p_2)+p_{2}^{\rho}p_{2\nu}\right]
\left[p_1^2D(p_1)\right]H_{\rho\alpha}(p_2,q) \nonumber \\
& - & \left[i\Delta^{(-1)\,\rho}_{\alpha}(q)+q^\rho q_\alpha\right]
\left[p_1^2D(p_1)\right]H_{\rho\nu}(q,p_2), \nonumber \\
p_2^\nu\g_{A_\alpha A_\mu A_\nu}(q,p_1,p_2) & = & 
\left[i\Delta^{(-1)\,\rho}_{\alpha}(q)+q^\rho q_\alpha\right]
\left[p_2^2D(p_2)\right]H_{\rho\mu}(q,p_1) \nonumber \\
& - & \left[i\Delta^{(-1)\,\rho}_{\mu}(p_1)+p_{1}^{\rho}p_{1\mu}\right]
\left[p_2^2D(p_2)\right]H_{\rho\alpha}(p_1,q), 
\label{STIbc}
\eea  where the  momenta $p_1^\mu$  and $p_2^\nu$  are now  related to
virtual  integration  momenta appearing  in  the  quantum loop.   This
construction    has    been     carried    out    at    one-loop    in
\cite{Cornwall:1989gv}   where   only   the  tree-level   version   of
Eq.(\ref{sti3gv}), namely Eq.(\ref{sti3gv0}),  has been invoked.  Here
we  present for  the first  time the  two-loop generalization  of this
construction, employing the one-loop version of Eq.(\ref{sti3gv}).  In
addition  to proving  that the  one-loop construction  can in  fact be
generalized to  two-loops, a non-trivial  result in its own  right, we
present  a  different  but  equivalent   point  of  view  to  that  of
\cite{Cornwall:1982zr}, motivated by the  BV formalism in general, and
the BQIs presented  in Section~\ref{sec:one} in particular.   
The novel ingredient
we  present  here is  the  following:  The  essential feature  of  the
intrinsic  PT construction  is to  arrive at  the desired  object, for
example the effective gluon  self-energy by discarding pieces from the
conventional self-energy.  The terms  discarded originate from the RHS
of Eq.(\ref{sti3gv}),  and, according  to the discussion  presented in
Section~\ref{sec:one}, 
they are  all precisely  known in  terms of  physical and
unphysical  Green's  functions, appearing  in  the  theory.  Then,  by
virtue  of  identifications such  as  those  given in  Eqs.\r{idena},
\r{Idb},  and~\r{HvsG}, one  can  directly compare  the
result obtained by the intrinsic PT procedure to the corresponding BFM
quantity  (at $\xi_Q=1$), employing  the BQI  of Eq.(\ref{bvsqtwopf}).
We emphasize that at no point do we use the BQIs or the BV formalism in
general  in arriving  at the  intrinsic PT  result; the  BQIs  are only
a-posteriori invoked,  at the  end of the  PT procedure,  because they
greatly facilitate the  comparison with the BFM result.   Last but not
least, the  two-loop construction presented  here, provides additional
evidence    supporting    the     point    of    view    adopted    in
\cite{Papavassiliou:1999az,Papavassiliou:1999bb},  namely that no
internal vertex  must be 
rearranged, and  no pieces must be  therefore discarded as  a result of
this rearrangement.  The  reason is simply that one  needs to maintain
the {\it full} one-loop three-gluon  vertex, on which the momenta will
act in  order for  Eq.(\ref{STIbc}) to be  triggered; instead,  if one
were    to   remove    pieces   by    modifying    internal   vertices
\cite{Watson:1998vv}, one would  invariably distort the aforementioned
STI.

We start by reviewing the one-loop intrinsic PT construction, 
beginning again, without loss of generality, in the renormalizable
Feynman gauge.
Consider the one-loop gluon two-point function
\be
\g^{(1)}_{A_\alpha A_\beta}(q)
=\frac12\int_{L_1}J(q,k)L_{\alpha\beta}(q,k),
\ee
where, after symmetrizing the ghost loop,
\be
L_{\alpha\beta}(q,k)=\Gamma_{\alpha \mu \nu}^{(0)}(q,k,-k-q)
\Gamma_{\beta}^{(0)\,\mu\nu}(q,k,-k-q)-k_\alpha 
\left(k+q\right)_\beta-k_\beta\left(k+q\right)_\alpha.
\ee
In the absence of longitudinal momenta coming from 
internal gluon propagators (since we work in the Feynman gauge),
the only momenta that can trigger an STI come from the 
three-gluon vertices.
We then carry out the PT decomposition 
of Eq.(\ref{decomp}) on {\it both} the three-gluon
vertices appearing at the two ends of the diagram, {\it i.e.}, we write
\bea
\Gamma^{(0)}_{\alpha \mu \nu}\Gamma^{(0)\,\mu \nu}_{\beta}  &=&
[\Gamma^{{\rm F}}_{\alpha \mu \nu} 
+ \Gamma^{{\rm P}}_{\alpha \mu \nu}]
[\Gamma^{{\rm F}\,\mu \nu}_{\beta} +
\Gamma^{{\rm P}\,\mu\nu}_{\beta}]
\nonumber\\
&=&
\Gamma^{{\rm F}}_{\alpha \mu \nu}
\Gamma^{{\rm F}\,\mu \nu}_{\beta}+
\Gamma^{{\rm P}}_{\alpha \mu \nu}
\Gamma^{(0)\,\mu \nu}_{\beta}+
\Gamma^{(0)}_{\alpha \mu \nu}
\Gamma^{{\rm P}\,\mu \nu}_{\beta} 
- 
\Gamma^{{\rm P}}_{\alpha \mu \nu}
\Gamma^{{\rm P}\,\mu\nu}_{\beta}.
\label{INPTDEC}
\eea
Of the four terms of the equation above, 
the first and last are left untouched;
for the second and third terms, using the 
three-gluon vertex STI of Eq.\r{STIbc}, we find
\bea
\Gamma^{{\rm P}}_{\alpha \mu \nu}
\Gamma^{(0)\,\mu\nu}_{\beta}+
\Gamma^{(0)}_{\alpha \mu \nu}
\Gamma^{{\rm P}\,\mu\nu}_{\beta} & = &
-4iq^2P_{\alpha\rho}(q)H^{(0)}_{\rho\beta}(q,-k-q)
+2ik^2P_{\alpha\rho}(k)H^{(0)}_{\rho\beta}(k,q) \nonumber \\
& + & 2i(k+q)^2P_{\alpha\rho}(k+q)H^{(0)}_{\rho\beta}(-k-q,q) \nonumber \\
& = & -4q^2P_{\alpha\beta}(q)+2k^2P_{\alpha\beta}(k)
+2(k+q)^2P_{\alpha\beta}(k+q).
\label{res1}
\eea
where (after factoring out the coupling constant $g$) 
we have used that $H^{(0)}_{\alpha\beta} =-ig_{\alpha\beta}$. 
The first term on the RHS, to be denoted by 
$\Pi_{\alpha\beta}^{{\rm IP}\,(1)}(q)$,  where the superscript
``IP'' stands for ``intrinsic pinch'', 
 is to be discarded from the gluon self-energy. Thus,
the 1PI one-loop intrinsic PT gluon self-energy,
to be denoted as before by ${\widehat\g}^{(1)}_{A_\alpha A_\beta}(q)$,
is defined as 
\be
{\widehat\g}^{(1)}_{A_\alpha A_\beta}(q) =
\g^{(1)}_{A_\alpha A_\beta}(q) - \Pi_{\alpha\beta}^{{\rm IP}\,(1)}(q) 
\label{1lPTtwopfIP}
\ee
Notice that $\Pi_{\alpha\beta}^{{\rm IP}\,(1)}(q)$ has precisely the form
\be
\Pi_{\alpha\beta}^{{\rm IP}\,(1)}(q) = 
\frac12\left[-4q^2P_{\alpha\beta}(q)\right]\int_{L_1}J(q,k)=
-\Pi_{\alpha\beta}^{{\rm P}\,(1)}(q),
\label{intPT}
\ee
so that dropping this term in Eq.\r{res1} has the same effect of canceling it
with the $S$-matrix PT.
At this point in the original construction of 
\cite{Cornwall:1989gv} the first and last terms on the RHS of
Eq.(\ref{INPTDEC}) were combined with the second and third term
on the RHS of  Eq.(\ref{res1}), in order to show that, after elementary
algebraic manipulations, 
${\widehat\g}^{(1)}_{A_\alpha A_\beta}(q)$ assumes the 
form 
\be
{\widehat\g}^{(1)}_{A_\alpha A_\beta}(q)
= \frac{1}{2} \int_{L_1} J_1(q,k)
\widehat{L}_{\alpha\beta}(q,k) \, ,
\label{PTprop2}
\ee
with
\be
\widehat{L}_{\alpha\beta}(q,k) \equiv 
\Gamma_{F\alpha}^{(0)\sigma\rho}(q,k,-k-q)
\Gamma_{F\beta\sigma\rho}^{(0)}(q,k,-k-q) - 
2 (2k+q)_{\alpha}(2k+q)_{\beta} \, .
\label{Tense2}
\ee
As was realized a few years later \cite{Denner:1994nn},   this last expression
of  ${\widehat\g}^{(1)}_{A_\alpha A_\beta}(q)$ coincides  with
$\g^{(1)}_{\widehat A_\alpha \widehat A_\beta}(q)$.  Notice however that, in
view of the BQI of  Eq.(\ref{bvsqtwopf}), this last identification is more
immediate, in the sense that  no further manipulation of the answer is needed:
the difference  between ${\widehat\g}^{(1)}_{A_\alpha A_\beta}(q)$ and
$\g^{(1)}_{A_\alpha A_\beta}(q)$ is the same as  the difference between 
$\g^{(1)}_{\widehat A_\alpha \widehat A_\beta}(q)$ and $\g^{(1)}_{A_\alpha
A_\beta}(q)$, as given by the BQI. Thus, once ${\widehat\g}^{(1)}_{A_\alpha
A_\beta}(q)$ has been constructed the BQI serves as a short-cut for relating it
to  $\g^{(1)}_{\widehat A_\alpha \widehat A_\beta}(q)$. Even though at 
one-loop the amount of algebra thusly saved is insignificant, at two-loops and
beyond the use of the BQI constitutes a definite technical advantage. 

Let us conclude the one-loop analysis by introducing a short-hand notation  for
the intrinsic PT construction, that will be useful in its  two-loop
generalization that will follow. We begin by writing the diagram of
Fig.\ref{fig2}a, suppressing all Lorentz and color indices, as
\be
(\ga\ga)=(\gf\gf)+(\gp\ga)+(\ga\gp)-(\gp\gp).
\ee
Then using the tree-level STI of Eq.\r{STIbc}, we write
\bea
(\gp\ga)&=&-iV\de +2L, \nonumber \\
(\ga\gp)&=&-i\de V+2L,
\eea
so that
\be
(\ga\ga)=-2i\de V+(\gf\gf)+4L-(\gp\gp).
\ee
The quantities $A$ and $L$, can be read off directly from Eq.\r{res1}, and are
equal to
\bea
V_{\alpha\beta}(q)&=& -2P_{\alpha\beta}(q) \int_{L_1}J(q,k),\nonumber\\
L_{\alpha\beta}(q)&=&
\frac12\int_{L_1}J(q,k)\left[k^2P_{\alpha\beta}(k)+(k+q)^2P_{\alpha\beta}(k+q)
\right].
\eea
Moreover one has the well known one-loop PT result
\be
(GG)+2L-\frac12(\gp\gp)=(\widehat G\widehat G),
\label{1lres}
\ee
where $(GG)$ represents the ghost diagram of Fig.\ref{fig2}b, and $(\widehat
G\widehat G)$ is the corresponding diagram with background ghost circulating in
the loop.
This will finally furnish the result
\bea
\g_{AA}^{(1)}&=&\frac12(\ga\ga)+(GG)\nonumber \\
&=&-i\de V+\frac12(\gf\gf)+(\widehat G\widehat G) \nonumber \\
&=&-\Pi^{{\rm P}\,(1)}+\g_{\widehat A\widehat A}^{(1)}\,.
\eea

\begin{figure}[!t]
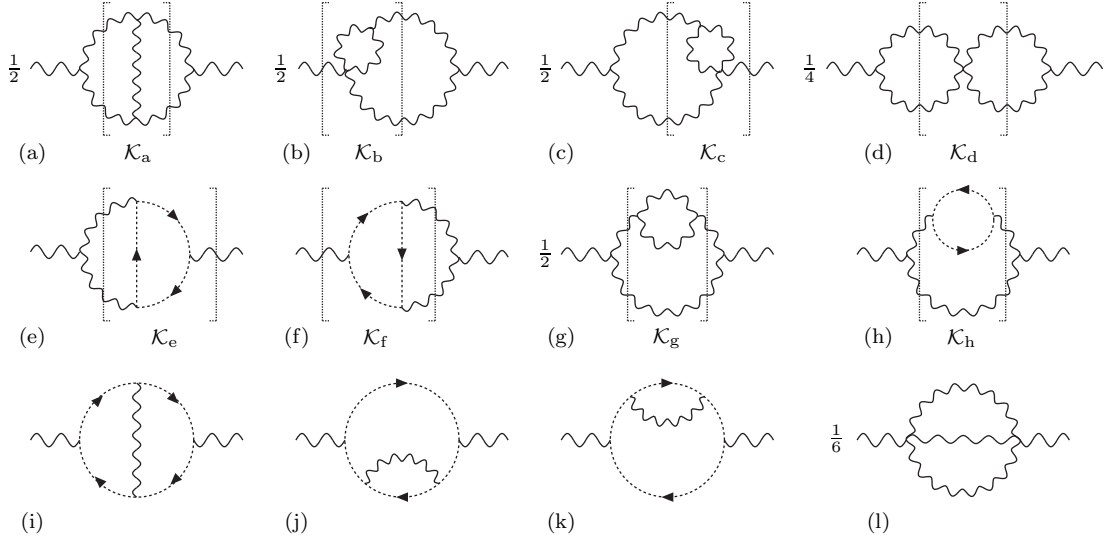

\bce
\bpi(0,200)(0,-155)

\Photon(-210,0)(-191.5,0){2.5}{2}
\PhotonArc(-170,0)(20,-6,354){1.5}{18}
\Photon(-170,-18.5)(-170,18.5){1.5}{6}
\Photon(-148.5,0)(-130,0){-2.5}{2}
\DashLine(-182.5,-25)(-182.5,25){0.5}
\DashLine(-182.5,25)(-180,25){0.5}
\DashLine(-182.5,-25)(-180,-25){0.5}
\DashLine(-157.5,-25)(-157.5,25){0.5}
\DashLine(-157.5,25)(-160,25){0.5}
\DashLine(-157.5,-25)(-160,-25){0.5}
\Text(-170,-32)[c]{$\scriptstyle {\cal K}_{\rm a}$}

\Photon(-109,0)(-91,0){2.5}{2}
\PhotonArc(-70,0)(20,181,489){1.5}{16}
\PhotonArc(-86.3,7.4)(8,2,362){1.5}{8}
\Photon(-48.5,0)(-30,0){-2.5}{2}
\DashLine(-70,-25)(-70,25){0.5}
\DashLine(-70,25)(-72.5,25){0.5}
\DashLine(-70,-25)(-72.5,-25){0.5}
\DashLine(-100,-25)(-100,25){0.5}
\DashLine(-100,25)(-97.5,25){0.5}
\DashLine(-100,-25)(-97.5,-25){0.5}
\Text(-82.5,-32)[c]{$\scriptstyle {\cal K}_{\rm b}$}

\Photon(-10,0)(8.5,0){2.5}{2}
\PhotonArc(30,0)(20,51.5,358){1.5}{15}
\PhotonArc(46.2,7.5)(8,12,372){1.5}{8}
\Photon(50.5,0)(70,0){-2.5}{2}
\DashLine(30,-25)(30,25){0.5}
\DashLine(30,-25)(32.5,-25){0.5}
\DashLine(30,25)(32.5,25){0.5}
\DashLine(61,-25)(61,25){0.5}
\DashLine(61,25)(58.5,25){0.5}
\DashLine(61,-25)(58.5,-25){0.5}
\Text(47.5,-32)[c]{$\scriptstyle {\cal K}_{\rm c}$}

\Photon(90,0)(108.5,0){2.5}{2}
\PhotonArc(125,0)(15,-6,354){1.5}{14}
\PhotonArc(158,0)(15,-6,354){1.5}{14}
\Photon(174.5,0)(194,0){-2.5}{2}
\DashLine(125,-25)(125,25){0.5}
\DashLine(125,25)(127.5,25){0.5}
\DashLine(125,-25)(127.5,-25){0.5}
\DashLine(158,-25)(158,25){0.5}
\DashLine(158,25)(155.5,25){0.5}
\DashLine(158,-25)(155.5,-25){0.5}
\Text(142.5,-32)[c]{$\scriptstyle {\cal K}_{\rm d}$}

\Photon(-210,-69)(-191.5,-69){2.5}{2}
\PhotonArc(-170,-70)(20,90,270){1.5}{9}
\DashCArc(-170,-70)(20,270,90){1}
\DashArrowLine(-170,-90)(-170,-50){1}
\Photon(-150,-70)(-130,-70){-2.5}{2}
\DashArrowLine(-156,-55.5)(-155.5,-56){1}
\DashArrowLine(-155.5,-84)(-156,-84.5){1}
\DashLine(-182.5,-45)(-182.5,-95){0.5}
\DashLine(-182.5,-95)(-180,-95){0.5}
\DashLine(-182.5,-45)(-180,-45){0.5}
\DashLine(-140,-45)(-140,-95){0.5}
\DashLine(-140,-95)(-142.5,-95){0.5}
\DashLine(-140,-45)(-142.5,-45){0.5}
\Text(-160,-102)[c]{$\scriptstyle {\cal K}_{\rm e}$}

\Photon(-110,-70)(-90,-70){2.5}{2}
\PhotonArc(-70,-70)(20,270,90){1.5}{9}
\DashCArc(-70,-70)(20,90,270){1}
\DashArrowLine(-70,-50)(-70,-90){1}
\Photon(-48.5,-71)(-30,-71){-2.5}{2}
\DashArrowLine(-84.5,-56)(-84,-55.5){1}
\DashArrowLine(-84,-84.5)(-84.5,-84){1}
\DashLine(-57.5,-95)(-57.5,-45){0.5}
\DashLine(-57.5,-45)(-60,-45){0.5}
\DashLine(-57.5,-95)(-60,-95){0.5}
\DashLine(-100,-45)(-100,-95){0.5}
\DashLine(-100,-95)(-97.5,-95){0.5}
\DashLine(-100,-45)(-97.5,-45){0.5}
\Text(-80,-102)[c]{$\scriptstyle {\cal K}_{\rm f}$}

\Photon(-10,-70)(9,-70){2.5}{2}
\PhotonArc(30,-72)(20,125,55){1.5}{13.5}
\PhotonArc(30,-57)(10,0,360){1.5}{9}
\Photon(51.5,-70)(70,-70){-2.5}{2}
\DashLine(15,-45)(15,-95){0.5}
\DashLine(15,-95)(17.5,-95){0.5}
\DashLine(15,-45)(17.5,-45){0.5}
\DashLine(45,-45)(45,-95){0.5}
\DashLine(45,-95)(42.5,-95){0.5}
\DashLine(45,-45)(42.5,-45){0.5}
\Text(30,-102)[c]{$\scriptstyle {\cal K}_{\rm g}$}

\Photon(101.5,-70)(120.5,-70){2.5}{2}
\PhotonArc(141.5,-72)(20,125,55){1.5}{13.5}
\DashCArc(141.5,-57)(11.5,0,360){1}
\Photon(163,-70)(181.5,-70){-2.5}{2}
\DashArrowLine(141,-68.5)(142,-68.5){1}
\DashArrowLine(142,-45.5)(141,-45.5){1}
\DashLine(125,-45)(125,-95){0.5}
\DashLine(125,-95)(127.5,-95){0.5}
\DashLine(125,-45)(127.5,-45){0.5}
\DashLine(157.5,-45)(157.5,-95){0.5}
\DashLine(157.5,-95)(155,-95){0.5}
\DashLine(157.5,-45)(155,-45){0.5}
\Text(141.5,-102)[c]{$\scriptstyle {\cal K}_{\rm h}$}

\Photon(-210,-140)(-191.5,-140){2.5}{2}
\DashCArc(-170,-140)(21.5,0,360){1}
\Photon(-170,-161.5)(-170,-118.5){1.5}{6}
\Photon(-148.5,-140)(-130,-140){-2.5}{2}
\DashArrowLine(-156,-124.5)(-155.5,-125){1}
\DashArrowLine(-155.5,-155)(-156,-155.5){1}
\DashArrowLine(-184.5,-125)(-184,-124.5){1}
\DashArrowLine(-184,-155.5)(-184.5,-155){1}

\Photon(-110,-140)(-91.5,-140){2.5}{2}
\DashCArc(-70,-140)(21.5,0,360){1}
\Photon(-48.5,-140)(-30,-140){-2.5}{2}
\DashArrowLine(-69.5,-161.5)(-70.5,-161.5){1}
\DashArrowLine(-70.5,-118.5)(-69.5,-118.5){1}
\PhotonArc(-70,-161.5)(15,20,160){-1.5}{6.5}

\Photon(-10,-140)(8.5,-140){2.5}{2}
\DashCArc(30,-140)(21.5,0,360){1}
\Photon(51.5,-140)(70,-140){-2.5}{2}
\DashArrowLine(30.5,-161.5)(29.5,-161.5){1}
\DashArrowLine(29.5,-118.5)(30.5,-118.5){1}
\PhotonArc(30,-118.5)(15,200,340){-1.5}{6.5}

\Photon(101.5,-140)(120,-140){2.5}{2}
\PhotonArc(141.5,-140)(20,-6,354){1.5}{18}
\Photon(120,-139.5)(163,-140){1.5}{4.5}
\Photon(163,-140)(181.5,-140){-2.5}{2}

\Text(-214,0)[r]{${\scriptstyle{\frac12}}$}
\Text(-113,0)[r]{${\scriptstyle{\frac12}}$}
\Text(-13,0)[r]{${\scriptstyle{\frac12}}$}
\Text(87,0)[r]{${\scriptstyle{\frac14}}$}
\Text(-13,-70)[r]{${\scriptstyle{\frac12}}$}
\Text(97,-140)[r]{${\scriptstyle{\frac16}}$}

\Text(-210,-32)[c]{\scriptsize{(a)}}
\Text(-110,-32)[c]{\scriptsize{(b)}}
\Text(-10,-32)[c]{\scriptsize{(c)}}
\Text(110,-32)[c]{\scriptsize{(d)}}

\Text(-210,-102)[c]{\scriptsize{(e)}}
\Text(-110,-102)[c]{\scriptsize{(f)}}
\Text(-10,-102)[c]{\scriptsize{(g)}}
\Text(110,-102)[c]{\scriptsize{(h)}}

\Text(-210,-172)[c]{\scriptsize{(i)}}
\Text(-110,-172)[c]{\scriptsize{(j)}}
\Text(-10,-172)[c]{\scriptsize{(k)}}
\Text(110,-172)[c]{\scriptsize{(l)}}
\epi
\ece
\caption{\label{figx0} 
The Feynman diagrams, together with their statistical weights and the
associated kernels, 
contributing to the conventional two-loop gluon
two-point function $\g_{A_\alpha A_\beta}^{(2)}(q)$ in the $R_\xi$ gauges.}
\end{figure}

We will now generalize the intrinsic PT construction presented above, to
two-loops.  The 1PI Feynman diagrams contributing  to the conventional two-loop
gluon self-energy in the $R_\xi$ gauges are represented in Fig.\ref{figx0}.  
They can be separated into three distinct sets: ({\it i}) the set of 
diagrams that
have two external (tree-level) three-gluon vertices, and thus can be written
schematically (suppressing Lorentz indices) as  $\Gamma^{(0)}[{\cal
K}]\Gamma^{(0)}$, where ${\cal K}$ is some kernel;  to this set belong diagrams
(a), (d), (g) and (h). ({\it ii})  the set of diagrams with  only one external
(tree-level) three-gluon vertex,  and thus can be written  as
$\Gamma^{(0)}[{\cal K}]$ or $[{\cal K}]\Gamma^{(0)}$; this set is composed by
the diagrams (b), (c), (e) and (f). ({\it iii}) All remaining
diagrams, containing no external three-gluon vertices. 

At this point we make the following observation: if one were to carry out  the
decomposition Eq.(\ref{INPTDEC}) to the pair of external  vertices appearing in
the diagrams of the set ({\it i}), and the decomposition of
Eq.(\ref{decomp}) 
to the
external vertex appearing in the diagrams of the set ({\it ii}), 
after a judicious
rearrangement of terms,  the longitudinal terms $p_{1}^{\mu}$ and $p_{2}^{\nu}$
stemming from  $\Gamma^{{\rm P}}_{\alpha \mu \nu}(q,p_1,p_2)$ 
and/or $\Gamma^{{\rm P}\,\mu\nu}_{\beta}(q,p_1,p_2)$  
would be triggering the one-loop version of
Eq.(\ref{STIbc}), just as in the one-loop case one has been triggering the 
tree-level version of Eq.(\ref{STIbc}). The only exception are of course
diagrams (g) and (h), where the STI triggered is still  the  tree-level version
of Eq.(\ref{STIbc}). Therefore,  the straightforward generalization of the
intrinsic PT to two-loops would amount to  isolating from the two-loop
diagrams the terms of the STI  of Eq.(\ref{STIbc}) that are proportional to 
$[\Delta^{(-1)\,\rho}_{\alpha}(q)]^{(n)}$, with  $n=0,1$; we will
denote such contributions by  $\Pi_{\alpha\beta}^{{\rm
IP}\,(2)}(q)$. Thus the 1PI diagrams contributing to the two-loop
gluon self-energy can be cast in the form
\be
\g_{A_\alpha A_\beta}^{(2)}(q)=G^{(2)}_{A_\alpha A_\beta}(q)+
\Pi_{\alpha\beta}^{{\rm IP}\,(2)}(q).
\label{ip1}
\ee 

\begin{figure}[!t]
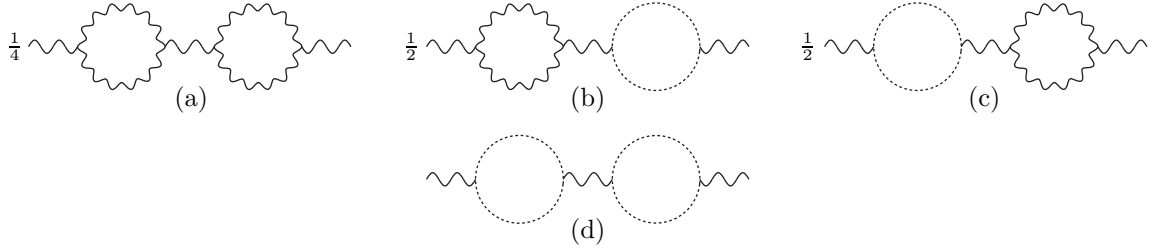

\bce
\bpi(0,80)(0,-60)

\Photon(-210,0)(-191.5,0){2.5}{2}
\PhotonArc(-175,0)(15,-6,354){1.5}{14}
\Photon(-158.5,0)(-140,0){2.5}{2}
\PhotonArc(-123.5,0)(15,-6,354){1.5}{14}
\Photon(-107,0)(-88.5,0){-2.5}{2}

\Photon(-60,0)(-41.5,0){2.5}{2}
\PhotonArc(-25,0)(15,-6,354){1.5}{14}
\Photon(-8.5,0)(10,0){2.5}{2}
\DashCArc(26.5,0)(16.5,-6,354){1}
\Photon(43,0)(61.5,0){-2.5}{2}

\Photon(90,0)(108.5,0){2.5}{2}
\DashCArc(125,0)(16.5,-6,354){1}
\Photon(141.5,0)(160,0){2.5}{2}
\PhotonArc(176.5,0)(15,-6,354){1.5}{14}
\Photon(193,0)(211.5,0){-2.5}{2}

\Photon(-60,-50)(-41.5,-50){2.5}{2}
\DashCArc(-25,-50)(16.55,-6,354){1}
\Photon(-8.5,-50)(10,-50){2.5}{2}
\DashCArc(26.5,-50)(16.5,-6,354){1}
\Photon(43,-50)(61.5,-50){-2.5}{2}

\Text(-213,0)[r]{$\scriptstyle{\frac14}$}
\Text(-63,0)[r]{$\scriptstyle{\frac12}$}
\Text(87,0)[r]{$\scriptstyle{\frac12}$}

\Text(-149.2,-20)[c]{\footnotesize{(a)}}
\Text(0.8,-20)[c]{\footnotesize{(b)}}
\Text(150.8,-20)[c]{\footnotesize{(c)}}
\Text(0.8,-70)[c]{\footnotesize{(d)}}
\epi
\ece

\caption{\label{fig1PR}
The two-loop 1PR strings 
(together with their statistical weights)
in the $R_\xi$ gauges.}

\end{figure}

Notice however that   
the 1PR set of one-loop self-energy diagrams, {\it i.e.}, the strings shown in
Fig.\ref{fig1PR}, 
must also be rearranged following the intrinsic PT procedure, and be
converted into the equivalent string involving PT one-loop
self-energies (which are known objects from the one-loop results). 
As we will  see in detail in what follows,
this treatment of the 1PR strings will
give rise, in addition to the PT strings, 
to ({\it a}) a set of contributions which are proportional to the
inverse propagator of the external legs $\de(q)$, and ({\it b})
a set of contributions which is {\it
effectively} 1PI, and therefore  also belongs to the definition of the 1PI
two-loop PT gluon self-energy; we will denote these two sets of contributions
collectively by $S^{{\rm IP}\,(2)}_{\alpha\beta}(q)$. Thus the sum of
the 1PI and 1PR contributions to the conventional two-loop gluon
self-energy can be cast in the form
\bea
\g_{A_\alpha A_\beta}^{(2)}(q)
+ \g_{A_\alpha A_\rho}^{(1)}(q)d(q)\g_{A^\rho A_\beta}^{(1)}(q)
&=&G^{(2)}_{A_\alpha A_\beta}(q)+\widehat\g_{A_\alpha A_\rho}^{(1)}(q)d(q)
\widehat\g_{A^\rho A_\beta}^{(1)}(q)\nonumber \\
&+&
\Pi_{\alpha\beta}^{{\rm IP}\,(2)}(q)+S^{{\rm
IP}\,(2)}_{\alpha\beta}(q).
\label{ip2}
\eea

By definition of the intrinsic PT procedure, we will now discard from the above
expression all the terms which are proportional to the inverse
propagator of the external legs $\de(q)$, thus defining the quantity
\be
R^{{\rm
IP}\,(2)}_{\alpha\beta}(q)=\Pi'^{\,{\rm IP}\,(2)}_{\alpha\beta}(q)+S'^{\,{\rm
IP}\,(2)}_{\alpha\beta}(q),
\label{ip3}
\ee
where the primed functions are defined starting from the unprimed ones
appearing in Eq.\r{ip2} by discarding the aforementioned terms.

Thus, making use of Eqs.\r{ip1}, \r{ip2} and \r{ip3},
the 1PI two-loop intrinsic PT gluon
self-energy, to be denoted as before by ${\widehat\g}^{(2)}_{A_\alpha
A_\beta}(q)$, is defined as 
\bea
{\widehat\g}^{(2)}_{A_\alpha A_\beta}(q)&=& G^{(2)}_{A_\alpha
A_\beta}(q)+R^{{\rm IP}\,(2)}_{\alpha\beta}(q)\nonumber \\
&=&\g^{(2)}_{A_\alpha A_\beta}(q) - \Pi_{\alpha\beta}^{{\rm IP}\,(2)}(q) +
R^{{\rm IP}\,(2)}_{\alpha\beta}(q).
\label{2lPTtwopfIP}
\eea
Of course the quantities 
${\widehat\g}^{(2)}_{A_\alpha A_\beta}(q)$ and 
$\g^{(2)}_{A_\alpha A_\beta}(q)$ appearing on the RHS
of Eqs.\r{2lPTtwopf} and~\r{2lPTtwopfIP} are
identical; however, the result of Eq.(\ref{intPT}) does not generalize 
beyond one-loop.
Thus, at two-loops
$-\Pi_{\alpha\beta}^{{\rm IP}\,(2)}(q) \neq 
\Pi_{\alpha\beta}^{{\rm P}\,(2)}(q)$
and $R_{\alpha\beta}^{{\rm IP}\,(2)}(q)
\neq R^{{\rm P}\,(2)}_{\alpha\beta}(q)$; however
\be
-\Pi_{\alpha\beta}^{{\rm IP}\,(2)}(q)+ R_{\alpha\beta}^{{\rm IP}\,(2)}(q)\equiv
\Pi_{\alpha\beta}^{{\rm P}\,(2)}(q)- R_{\alpha\beta}^{{\rm P}\,(2)}(q).
\label{ip4}
\ee

We next proceed to give the details of the construction of the
quantities $\Pi_{\alpha\beta}^{{\rm IP}\,(2)}(q)$ 
and~$R_{\alpha\beta}^{{\rm IP}\,(2)}(q)$ 
discussed above, starting from
the first one.

As a first step, we carry out the usual PT
decomposition of the three-gluon vertex to  
the graphs of set ({\it i}) and ({\it ii}). 
 Taking into account the statistical
factors, for diagrams (a) and (d) we then obtain 
\bea
\frac12\Gamma^{(0)}\left[{\cal K}_{\rm a}+{\scriptstyle{\frac12}}{\cal
K}_{\rm d}\right]\Gamma^{(0)} &=&
\frac12\Gamma^{{\rm F}}\left[{\cal K}_{\rm a}+
{\scriptstyle{\frac12}}{\cal K}_{\rm d}\right]\Gamma^{{\rm F}}+
\frac12\Gamma^{{\rm P}}
\left[{\cal K}_{\rm a}+{\scriptstyle{\frac12}}{\cal K}_{\rm d}\right]\Gamma^{(0)}
\nonumber \\
& + &
\frac12\Gamma^{(0)}\left[{\cal K}_{\rm a}+{\scriptstyle{\frac12}}{\cal
K}_{\rm d}\right]\Gamma^{{\rm P}}-
\frac12\Gamma^{{\rm P}}
\left[{\cal K}_{\rm a}+{\scriptstyle{\frac12}}{\cal
K}_{\rm d}\right]\Gamma^{{\rm P}}.
\label{dec2L}
\eea

As in the one-loop case, 
of the four terms appearing above the first and last term 
remain untouched, and constitute part of the answer. 
To the second and
third terms we add the $\Gamma^{{\rm P}}$ part of diagrams (c), (e)
and (b), (f) respectively to get
\bea
\left[{\rm (ii)+(c)+(e)}\right]^{{\rm P}}&=&
\frac12\Gamma^{{\rm P}}\left[{\cal K}_{\rm a}\Gamma^{(0)}+
{\scriptstyle{\frac12}}{\cal K}_{\rm d}\Gamma^{(0)}+{\cal K}_{\rm c}
+2{\cal K}_{\rm e}\right], \nonumber\\
\left[{\rm (iii)+(b)+(f)}\right]^{{\rm P}}&=&
\left[\Gamma^{(0)}{\cal K}_{\rm a}+
\Gamma^{(0)}{\scriptstyle{\frac12}}{\cal K}_{\rm d}+{\cal K}_{\rm b}
+2{\cal K}_{\rm f}\right]
\frac12\Gamma^{{\rm P}}.
\eea
It is then straightforward to see that the two contribution are actually equal,
and moreover that
\bce
\bpi(0,110)(210,-50)

\Text(220,50)[c]{${\cal K}_{\rm a}\Gamma^{(0)}+
{\scriptstyle{\frac12}}{\cal K}_{\rm d}\Gamma^{(0)}+{\cal K}_{\rm c}
+2{\cal K}_{\rm e}=$}

\Photon(10,20)(30,0){1.5}{6}
\Photon(10,-20)(30,0){-1.5}{6}
\Photon(30,0)(55,0){1.5}{5}
\Photon(18,-10)(18,10){1.5}{4}

\Photon(85,20)(105,0){1.5}{6}
\Photon(85,-20)(105,0){-1.5}{6}
\PhotonArc(113,0)(7,-8,352){1.5}{10}
\Photon(121.5,0)(136.5,0){1.5}{3}

\Photon(172.5,-20)(192.5,0){-1.5}{6}
\Photon(172,20)(181.5,10.5){1.5}{3}
\PhotonArc(186.5,5.5)(6,-8,352){1.5}{8}
\Photon(192.5,0)(207.5,0){-1.5}{3}

\Photon(242.5,20)(262.5,0){1.5}{6}
\Photon(242.5,-20)(252,-10.5){1.5}{3}
\PhotonArc(257,-6)(6,-8,352){1.5}{8}
\Photon(262,0)(277,0){1.5}{3}

\Photon(312.5,20)(322,10.5){1.5}{3}
\Photon(312.5,-20)(322,-10.5){1.5}{3}
\DashArrowLine(322,-10.5)(322,10.5){1}
\DashArrowLine(322,10.5)(332.5,0){1}
\DashArrowLine(332.5,0)(322,-10.5){1}
\Photon(332,0)(347,0){1.5}{3}

\Photon(382.5,20)(392,10.5){1.5}{3}
\Photon(382.5,-20)(392,-10.5){1.5}{3}
\DashArrowLine(392,10.5)(392,-10.5){1}
\DashArrowLine(392,-10.5)(402.5,0){1}
\DashArrowLine(402.5,0)(392,10.5){1}
\Photon(402,0)(417,0){1.5}{3}

\Text(60,-1)[l]{$+\ {\scriptstyle{\frac12}}$}
\Text(150,-1)[l]{$+\ {\scriptstyle{\frac12}}$}
\Text(220,-1)[l]{$+\ {\scriptstyle{\frac12}}$}
\Text(295,-1)[l]{$+$}
\Text(365,-1)[l]{$+$}

\Text(220,-50)[c]{$\equiv\ \g_{AAA}^{(1)}$.}

\epi
\ece
Thus, inserting back the Lorentz structure, we get the equation
\bea
[{\rm (ii)+(c)+(e)+(iii)+(b)+(f)}]^{{\rm P}}=\int_{L_1}\!
J(k,q)\Gamma^{{\rm P}}_{\alpha\mu\nu}(q,k,-k-q)
\g^{(1)}_{A_\beta A^\mu A^\nu}(q,k,-k-q). \nonumber \\
\label{intPT1}
\eea

For the remaining two first class diagrams (g) and (h), we  carry out the same
decomposition as for diagrams (a) and (d), concentrating again only on
the terms
\be
[{\rm (g)+(h)}]^{\rm P}=\Gamma^{{\rm P}}\left[
{\scriptstyle{\frac12}}{\cal K}_{\rm g}+{\cal K}_{\rm h}\right]\Gamma^{(0)}
+ \Gamma^{(0)}\left[
{\scriptstyle{\frac12}}{\cal K}_{\rm g}+{\cal K}_{\rm h}
\right]\Gamma^{{\rm P}}.
\ee
Next, one notices that the two contributions are actually equal,
and moreover that
\be
\bpi(0,40)(0,-15)

\Text(0,0)[r]{$\left[
{\scriptstyle{\frac12}}{\cal K}_{\rm g}+{\cal K}_{\rm h}\right]\Gamma^{(0)}
=\,$}

\Photon(10,20)(30,0){1.5}{6}
\Photon(10,-20)(30,0){-1.5}{6}
\Photon(30,0)(55,0){1.5}{5}
\GCirc(20,10){5}{0.8}

\epi
\ee
where the blob represent the one-loop correction to the gluon propagator.
Inserting back the Lorentz structure, we then get the equation
\bea
[{\rm (g)+(h)}]^{{\rm P}}=2\int_{L_1}\!J(k,q)\Gamma_{\alpha\mu\nu}^{{\rm P}}
(q,k,-k-q)
\g^{(1)}_{A^\mu A_\rho}(k)d(k)\Gamma_\beta^{(0)\,\rho\nu}
(q,k,-k-q).
\label{intPT2}
\eea

Eqs.\r{intPT1} and~\r{intPT2} will then be our starting point: 
from them, by
using the three-gluon vertex STI of Eq.\r{STIbc}, we will
isolate the 1PI parts that are proportional to the inverse propagator of the
external leg, and simply discard them. 

Let us start from Eq.\r{intPT1}. From Eq.\r{GFGP} and 
the one-loop version of Eq.\r{STIbc} we find
\bea
\Gamma^{{\rm P}}_{\alpha\mu\nu}(q,k,-k-q)
\g^{(1)}_{A_\beta A^\mu A^\nu}(q,k,-k-q) & = &
2H^{(0)}_{\rho\alpha}(q,-k-q)\g^{(1)}_{A^\rho A_\beta}(q)\nonumber \\
&-& 2q^2H^{(0)}_{\rho\alpha}(q,-k-q)P^\rho_\beta(q)\left[k^2D^{(1)}(k)\right]
\nonumber \\
&-& 2iq^2H^{(1)}_{\rho\alpha}(q,-k-q)P^\rho_\beta(q)+\dots,
\label{Ell}
\eea
where the ellipses stands for terms that will be part of the two-loop
function $G^{(2)}_{A_\alpha A_\beta}(q)$.
The perturbative expansion of the function $H$, will give
\bce
\bpi(0,40)(20,-15)

\Text(-5,2)[r]{$H^{(1)}_{\rho\beta}(q,-k-q)=\,$}

\PhotonArc(30,0)(20,30,180){-1.5}{8.5}
\DashCArc(30,0)(20,180,330){1}
\Photon(30,-20)(30,18.5){1.5}{6}
\DashArrowLine(42.6,-15.5)(41,-16.7){1}
\DashArrowLine(14.5,-12)(13.8,-11){1}
\Vertex(10,0){1.8}

\DashCArc(115,0)(20,90,180){1}
\PhotonArc(115,0)(20,30,90){-1.5}{4}
\PhotonArc(115,0)(20,180,270){-1.5}{5}
\DashCArc(115,0)(20,270,330){1}
\DashArrowLine(115,-20)(115,20){1}

\DashArrowLine(128.6,-14.7)(127,-16.2){1}

\DashArrowLine(99.5,12)(98.8,11){1}
\Vertex(95,0){1.8}

\Text(75,0)[c]{$+$}
\Text(55,10)[c]{$\scriptstyle \beta$}
\Text(140,10)[c]{$\scriptstyle \beta$}
\Text(20,0)[r]{$\scriptstyle \rho$}
\Text(105,0)[r]{$\scriptstyle \rho$}

\epi
\ece
so that, using the Feynman rules given in Fig.\ref{fig0}, we find
\bea
& & \int_{L_1}\!J(q,k)\left[2H^{(0)}_{\rho\alpha}(q,-k-q)\g^{(1)}_{A^\rho
A_\beta}(q)\right]=  
-i\g^{(1)}_{A_\alpha A_\rho}(q)V^{{\rm P}\,(1)\,\rho}_\beta(q), \label{first}\\
& &\int_{L_1}\!J(q,k)\left\{-2q^2H^{(0)}_{\rho\alpha}(q,-k-q)
P^\rho_\beta(q)\left[k^2D^{(1)}(k)\right]\right\}=-q^2P_{\alpha\beta}(q)I_3,  
\label{second}\\
& &\int_{L_1}\!J(q,k)\left[-2iq^2H^{(1)}_{\rho\alpha}(q,-k-q)P^\rho_\beta(q)
\right]=q^2I_1 \big\{\big[ k_{\rho}g_{\alpha\sigma} \nonumber
\\ & &\hspace{5.37cm} 
+\,\Gamma_{\sigma\rho\alpha}^{(0)}(-k,-\ell,k+\ell)\big](\ell-q)^{\sigma}
\big\}P^\rho_\beta(q).
\label{third}
\eea
 
Next consider the lower order corrections of Eq.\r{intPT2}.
From Eq.\r{GFGP} and the fact that the gluon two-point
function is transverse at all orders, we find
\bea
& & \Gamma_{\alpha\mu\nu}^{{\rm P}}
(q,k,-k-q)
\g^{(1)}_{A^\mu A_\rho}(k)\Gamma_\beta^{(0)\,\rho\nu}
(q,k,-k-q)\nonumber \\
&  & \hspace{4.0cm}
=\,-(k+q)_\nu\g^{(1)}_{A_\alpha A_\rho}(k)
\Gamma_\beta^{(0)\,\rho\nu}(q,k,-k-q), 
\eea
so that using the tree-level version of the STI of Eq.\r{STIbc} and isolating 
the term which will be discarded, we have  
\be
2J(q,k)\g^{(1)}_{A_\alpha A_\rho}(k)d(k)
\left[-iq^2P^\sigma_\beta(q)H^{(0)}_{\sigma\rho}(q,k)\right]
=-q^2P^\rho_\beta(q)I_4L_{\alpha\rho}(\ell,k) .
\label{fourth}
\ee
Collecting the terms on the RHS of 
Eqs.\r{first}--\r{fourth} we finally obtain
\bea
\Pi_{\alpha\beta}^{{\rm IP}\,(2)}(q) &=&
-i\g^{(1)}_{A_\alpha A_\rho}(q)V^{{\rm P}\,(1)\,\rho}_\beta(q)
-q^2 P_{\alpha\beta}(q)I_3
-q^2P^\rho_\beta(q)I_4 L_{\alpha\rho}(\ell,k)
\nonumber \\
&+& q^2P_\alpha^\rho(q) I_1 \big[ k_{\rho}g_{\alpha\sigma} 
+\,\Gamma_{\sigma\rho\beta}^{(0)}(-k,-\ell,k+\ell)\big](\ell-q)^{\sigma}
 \big]
\label{PIIP}
\eea

Next we turn to  the contributions  coming from the the conversion of the
conventional two-loop 1PR strings
to PT 1PR strings, and
determine the quantity $S_{\alpha\beta}^{{\rm IP}\,(2)}(q)$.
Using the notation introduced in the one-loop case we find for the 
diagram of Fig.\ref{fig1PR}a the result
\bea
{(5\rm a)}&=& (\ga\ga)\di(\ga\ga) \nonumber \\
&=& (\gf\gf)\di(\gf\gf)-(\gf\gf)\di(\gp\gp)-(\gp\gp)\di(\gf\gf)+
(\gp\gp)\di(\gp\gp) \nonumber \\
&+& 4(\gf\gf)\di L +4L\di(\gf\gf)-4(\gp\gp)\di L -4L\di(\gp\gp)
+16L\di L  \nonumber \\
&-& i(\gf\gf)\di V\de-i\de V\di(\gf\gf)+i\de V\di(\gp\gp)+i(\gp\gp)\di V\de
\nonumber \\
&-&\de V\di V\de-4i\de V\di L-4i L\di V\de \nonumber \\
&+& V\di^{-1}V-iV(\ga\ga)-i(\ga\ga)V.
\label{5a}
\eea
Here we have made an explicit 
distinction between the {\it internal} propagator $d_{\rm
i}$ and the external ones $d$: The presence or absence of the former will
determine if the corresponding diagram has to be considered 1PR or 1PI
respectively.
For the remaining diagrams we then get
\bea
{(5\rm b)}&=&(\ga\ga)\di(GG) \nonumber \\
&=& (\gf\gf)\di(GG)-(\gp\gp)\di(GG)+4L\di(GG)-i\de
V\di(GG)-iV(GG),\quad
\qquad \nonumber\\
{(5\rm c)}&=&(GG)\di(\ga\ga) \nonumber \\
&=& (GG)\di(\gf\gf)-(GG)\di(\gp\gp)+4(GG)\di L-i(GG)\di V\de -i(GG)V.
\eea

We can then start collecting pieces. Recalling the statistical weight of each 
diagram of Fig.\ref{fig1PR}, and using the one-loop result 
of Eq.\r{1lres}, we find
\bea
\frac12(\gf\gf)\di(\widehat G \widehat G) &=&
\frac12(\gf\gf)\di\left[(GG)+2L-\frac12(\gp\gp)\right], \nonumber\\
\frac12(\widehat G \widehat G)\di(\gf\gf) &=&
\frac12\left[(GG)+2L-\frac12(\gp\gp)\right]\di(\gf\gf),\nonumber \\
(\widehat G \widehat G)\di(\widehat G \widehat G) &=&
(GG)\di(GG)+(GG)\di\left[2L -\frac12(\gp\gp)\right]+
\left[2L -\frac12(\gp\gp)\right]\di(GG) \nonumber \\
&+&4L\di L-L\di(\gp\gp)-(\gp\gp)\di L+
\frac14(\gp\gp)\di(\gp\gp).
\eea
These terms together with the first term of Eq.\r{5a}, will give the PT 1PR
string. For the genuine 1PI terms we have instead the following result
\be
\frac14 V\di^{-1}V-\frac14 iV(\ga\ga)-\frac14 i(\ga\ga)V-
\frac12 iV(GG)-\frac12 i(GG)V=\frac14 V\di^{-1}V-iV\g^{(1)}_{AA}, 
\ee
so that by adding to them the
remaining terms proportional to the {\it external} inverse propagator
$\de(q)$, we will get the quantity     
\bea
S^{{\rm IP}\,(2)}(q) &=& \frac14 V\di^{-1}V-iV\g^{(1)}_{AA} \nonumber \\
&-& i(\gf\gf)\di V\de-i\de V\di(\gf\gf)+i\de V\di(\gp\gp)+i(\gp\gp)\di V\de
\nonumber \\
&-&\de V\di V\de-4i\de V\di L-4i L\di V\de-i\de
V\di(GG)-i(GG)\di V\de. \nonumber \\
\label{SIP}
\eea

Then if, according to the intrinsic PT algorithm, we discard from
Eqs.\r{PIIP} and~\r{SIP} the terms  proportional to the external
inverse propagator, we find
\bea
\Pi'^{\,{\rm IP}\,(2)}_{\alpha\beta}(q)&=&-i\g_{A_\alpha
A_\rho}^{(1)}(q)V^{{\rm P}\,(1)\,\rho}_\beta(q), \nonumber \\
S'^{\,{\rm IP}\,(2)}_{\alpha\beta}(q)&=&-q^2I_2P_{\alpha\beta}(q)+
i\g_{A_\alpha A_\rho}^{(1)}(q)V^{{\rm P}\,(1)\,\rho}_\beta(q),
\eea
so that adding by parts the equations above we obtain
\be
R^{{\rm IP}\,(2)}_{\alpha\beta}(q)=-q^2I_2P_{\alpha\beta}(q).
\label{RIP}
\ee
Thus, finally, the quantity $-\Pi^{{\rm
IP}\,(2)}_{\alpha\beta}(q)+R^{{\rm IP}\,(2)}_{\alpha\beta}(q)$ will
provide precisely the expressions appearing in the two-loop version of
the relevant BQI, {\it i.e.}, Eq.\r{bvsqtwopf2l}; or, equivalently,
Eq.\r{ip4} is proved. Notice that if instead
of resorting to the BQI one were to attempt a direct comparison
of the answer to the two-loop BFM gluon self-energy, one would have to:
({\it i}) collect the pieces denoted by ellipses in  
Eq.(\ref{Ell}); ({\it ii}) add them to the first and fourth term of 
Eq.(\ref{dec2L}); ({\it iii}) add the  $R_{\alpha\beta}^{{\rm IP}\,(2)}(q)$
of Eq.(\ref{RIP})-- at that point we have the PT result of Eq.\r{2lPTtwopfIP}.
({\it vi}) To compare the answer to that of the BFM we need to algebraically
manipulate the result; most notably
one must recover the very
characteristic ghost structure emerging in the BFM framework,
and in particular the appearance of four-particle ghost-vertices.
This straightforward but laborious 
procedure of algebraically recovering from the PT answer 
all the individual Feynman diagrams appearing in the BFM
has been followed in the original two-loop
presentation \cite{Papavassiliou:1999az,Papavassiliou:1999bb}, 
in the context of the 
$S$-matrix PT. 

\section{\label{sec:five} Conclusions}

In this  paper we  have formulated for  the first  time the PT  not in
terms of the elementary WI  satisfied by the bare, tree-level vertices
of the theory, but instead in terms of the STI satisfied by the higher
order  (one-loop  and  higher)   vertices.   In  particular,  the  STI
satisfied  by the  one-loop three-gluon  vertex allows  one to  take a
first  step  towards  a  non-diagrammatic  implementation  of  the  PT
algorithm: instead of  manipulating {\it individual} Feynman diagrams,
entire sets of  such diagrams are treated at  once. In particular, the
pieces that are reassigned from  the vertices to the self-energies (or
vice-versa)  can be  collectively  identified with  the ghost  Green's
functions  appearing  in  the   STI;  these  ghost  Green's  functions
determine  the deviation  of the  STI  from the  naive, tree-level  WI.
In  order to  avoid possible  confusions we
emphasize that the STI are  employed for the subset of diagrams nested
inside  the  higher  order  loops, in  the  conventional  formulation;
however, the  final one- and  two-loop effective PT  Green's functions
obtained through  this procedure  do not satisfy STIs but  instead the
characteristic naive, QED-like WIs known from the earlier literature on
the subject \cite{Cornwall:1982zr,Cornwall:1989gv}.

For comparing the  PT results to those of the BFM,  we have employed a
set of  identities relating the conventional Green's  functions to the
corresponding ones  in the BFM; the  two sets are related  by means of
auxiliary  Green's   functions  built  from   background  sources  and
anti-fields,  which are  characteristic of  the BV  formalism  we have
employed  for arriving  at them.  It  turns out  that these  auxiliary
Green's  functions  are  connected  to  the  ghost  Green's  functions
appearing  in the STI  by Eq.(\ref{gpert1}).   It is  interesting that
even though they originate from entirely different formalisms, the two
sets  of  unphysical Green's  functions  are  related  by such  simple
expressions.  Quite  remarkably,   the  PT  exposes  these  underlying
relations,  which appear  to be  encoded, in  a non-manifest  and very
intricate way, into physical observables, such as $S$-matrix elements.

It is  worth reviewing  briefly  some of  the main  physical
application  of  the  PT  in  the context  of  QCD.   The  unambiguous
construction    \cite{Grunberg:1989xf,Grunberg:1992mp,Haussling:1998pp}
of            the           universal           (process-independent),
gauge-fixing-parameter-independent,     scale- and scheme-invariant
effective  charge is  of  significant interest  \cite{Cornwall:1982zr,
Cornwall:1989gv,Papavassiliou:1990zd,Watson:1997fg}.  This  PT
construction allows for the 
explicit identification of  the conformally-(in)variant subsets of QCD
graphs \cite{Brodsky:2000cr,Rathsman:2001xe,Brodsky:1983gc}.
This is of relevance in the field of renormalon
calculus, where one studies the onset of non-perturbative effects from
the  behaviour  near  the  QCD  mass-scale  of  appropriately  selected
infinite sub-sets of the perturbative series \cite{Peris:1997dq}.

The systematic study of the interface
between perturbative and non-perturbative QCD is a 
long-standing problem.
It has been advocated that  the non-perturbative QCD effects can
be reliably captured  at  an inclusive level  by means  of an infrared
finite  quantity, which  would  constitute    the  extension of    the
perturbative  QCD  running coupling  to  low energy  scales 
\cite{Dokshitzer:1996qm}.  
Early results \cite{Cornwall:1982zr} 
based on the study  of gauge-invariant
Schwinger-Dyson  equations involving  this  quantity suggest
that such a description can in fact  be derived from first principles.
According to this  analysis, the self-interaction  of gluons give rise
to a dynamical gluon mass, while preserving at the same time the local
gauge  symmetry  of   the theory.  The  presence   of  the gluon  mass
saturates the  running of the  QCD coupling; so, instead of increasing
indefinitely in the   infrared  as perturbation  theory predicts,   it
``freezes'' at a finite value 
\cite{Cornwall:1982zr,Cornwall:1989gv,Papavassiliou:1990zd};
for an interesting discussion on the phenomenological implications of 
the various ``freezing'' models and mechanisms 
available in the literature, see  \cite{Aguilar:2001zy}. 

Finally, as has been pointed out by Brodsky in a series of  
recent papers \cite{Brodsky:2001ha, Brodsky:2001wx, Brodsky:2001ww}, 
the PT effective charge   
can serve as the natural scheme for defining the coupling 
in the proposed ``event amplitude generators''
 based on the the light-cone formulation of QCD.  

It is interesting to extend the analysis  
presented here for the case of QCD to the more involved context 
of theories with spontaneous symmetry breaking, in general, and the
Elecroweak Sector of the Standard Model in particular.  
A detailed analysis \cite{Binosi:2002bs}
reveals that the BV formalism is 
particularly suited for accomplishing the two-loop generalization 
of the PT in the Elecroweak Sector, 
a task which, due to the proliferation of
Feynman diagrams and the non-transversality 
of the gauge boson self-energies, has been pending.

We believe that the methodology and the formal connections established in
this paper set up the stage for the formulation of the PT to all
orders in perturbation theory \cite{DBJP2}. 
It remains to be seen whether 
the PT will transcend its humble diagrammatic origins
and acquire the stature of a well-defined formal tool.

\begin{acknowledgments}
D.B. acknowledges useful correspondence with G.~Barnich 
and T.~Hurth. J.P. thanks P.A.~Grassi for various 
valuable discussions and communications. 
The work of D.B. is supported by the Ministerio de Ciencia y
Tecnolog\'\i a, Spain, under Grant BFM2001-0262, and 
the research of J.P. is supported by CICYT, Spain, under Grant AEN-99/0692.  
\end{acknowledgments}

\bibliography{ptallbib.bib}

\begin{thebibliography}{60}
\bibitem[{Ack()}]{Ack}
\bibinfo{note}{We are greatful to P. A. Grassi for pointing this out and for
  kindly furnishing to us the details for their derivation}.
\end{thebibliography}

\end{document}